\documentclass[twocolumn]{aastex701}

\usepackage{afterpage}

\usepackage{booktabs} % For \toprule, \midrule and \bottomrule
\let\tablenum\relax 
\usepackage{siunitx} % Formats the units and values
\usepackage{amsmath,amstext,color}
\usepackage{gensymb}
\usepackage{natbib}
\usepackage{hyperref}
\usepackage{longtable}
\usepackage{threeparttablex}
\usepackage{booktabs}
\usepackage{float} 

\usepackage{color}
\usepackage{graphicx}
\usepackage{amsmath}
\usepackage{amssymb}

\def\rr#1{#1}

\usepackage[]{appendix}
\begin{document}
\title{A Multi-Wavelength View of the First Type Ic-BL Supernova with an Einstein Probe X-ray Shock Breakout}
\correspondingauthor{Jillian Rastinejad}
\email{jcrastin@umd.edu}

\shorttitle{EP\,260321a/SN\,2026gzf}
\shortauthors{Rastinejad et al.}
\author[0000-0002-9267-6213]{Jillian C.~Rastinejad}
\altaffiliation{NASA Einstein Fellow}
\affiliation{Department of Astronomy, University of Maryland, College Park, MD~20742, USA}
\email{jcrastin@umd.edu}

\author[0000-0002-6428-2700]{Gokul Srinivasaragavan}
\affiliation{Department of Astronomy, University of Maryland, College Park, MD~20742, USA}
\affiliation{Joint Space-Science Institute, University of Maryland, College Park, MD 20742, USA}
\affiliation{Astrophysics Science Division, NASA Goddard Space Flight Center, 8800 Greenbelt Rd, Greenbelt, MD 20771, USA}
\affiliation{Division of Physics, Mathematics, and Astronomy, California Institute of Technology, Pasadena, CA 91125, USA}
\email{gsriniv2@umd.edu}

\author[0000-0003-2700-1030]{Nikhil Sarin}
\affiliation{Kavli Institute for Cosmology, University of Cambridge, Madingley Road, CB3 0HA, UK}
\affiliation{Institute of Astronomy, University of Cambridge, Madingley Road, CB3 0HA, UK}
\email{nsarin.astro@gmail.com}

\author[0009-0007-1842-7028]{Tanner O'Dwyer}
\affiliation{William H. Miller III Department of Physics and Astronomy, Johns Hopkins University, Baltimore, Maryland 21218, USA}
\email{todwyer1@jh.edu}

\author[0000-0003-1673-970X]{S.~Bradley Cenko} 
\affiliation{Joint Space-Science Institute, University of Maryland, College Park, MD 20742, USA} 
\affiliation{Astrophysics Science Division, NASA Goddard Space Flight Center, 8800 Greenbelt Rd, Greenbelt, MD 20771, USA}  
\affiliation{Department of Physics, George Washington University, 725 21st St NW, Washington, DC, 20052, USA}
\email{brad.cenko@nasa.gov}

\author[0000-0002-9415-3766]{James K.\ Leung}
\affiliation{David A. Dunlap Department of Astronomy \& Astrophysics, University of Toronto, 50 St. George St, Toronto, ON M5S 3H4, Canada}
\affiliation{Dunlap Institute for Astronomy and Astrophysics, University of Toronto, 50 St. George Street, Toronto, ON M5S 3H4, Canada}
\affiliation{Racah Institute of Physics, The Hebrew University of Jerusalem, Jerusalem 91904, Israel}
\email{jamesk.leung@utoronto.ca}

\author[0000-0002-2028-9329]{Anya E. Nugent}
\affiliation{Center for Astrophysics\:$|$\:Harvard \& Smithsonian, 60 Garden St. Cambridge, MA 02138, USA}
\email{anya.nugent@cfa.harvard.edu}

\author[0000-0001-8472-1996]{Daniel~A.~Perley}
\affiliation{Astrophysics Research Institute, Liverpool John Moores University, 146 Brownlow Hill, Liverpool L3 5RF, UK}
\email{d.a.perley@ljmu.ac.uk}

\author[0000-0001-9915-8147]{Genevieve~Schroeder}
\affiliation{Department of Astronomy, Cornell University, Ithaca, NY 14853, USA}
\email{gms279@cornell.edu}

\author[0000-0003-3768-7515]{Shreya Anand}
\altaffiliation{LSST-DA Catalyst Postdoctoral Fellow}
\affiliation{Kavli Institute for Particle Astrophysics and Cosmology, Stanford University, 452 Lomita Mall, Stanford, CA 94305, USA}
\affiliation{Department of Physics, Stanford University, 382 Via Pueblo Mall, Stanford, CA 94305, USA}
\email{sanand08@stanford.edu}

\author[0000-0002-2184-6430]{Tom\'as Ahumada}
\affiliation{Cerro Tololo Inter-American Observatory/NSF NOIRLab, Casilla 603, La Serena, Chile}
\email{tomas.ahumada@noirlab.edu}

\author[0000-0002-8977-1498]{Igor Andreoni} \affiliation{University of North Carolina at Chapel Hill, 120 E. Cameron Ave., Chapel Hill, NC 27514, USA}
\email{igor.andreoni@unc.edu}

\author[0009-0008-2714-2507]{Aleksandra Bochenek}
\affiliation{Astrophysics Research Institute, Liverpool John Moores University, 146 Brownlow Hill, Liverpool L3 5RF, UK}
\email{a.m.bochenek@2023.ljmu.ac.uk}

\author[0000-0001-8104-3536]{Alessandra Corsi}
\affiliation{William H. Miller III Department of Physics and Astronomy, Johns Hopkins University, Baltimore, Maryland 21218, USA}
\email{acorsi2@jh.edu}

\author[0000-0002-4223-103X]{Christoffer Fremling}
\affiliation{Division of Physics, Mathematics, and Astronomy, California Institute of Technology, Pasadena, CA 91125, USA}
\affiliation{Caltech Optical Observatories, California Institute of Technology, Pasadena, CA 91125, USA}
\email{fremling@caltech.edu}

\author[0000-0002-9017-3567]{Anna Y. Q. Ho}
\affiliation{Department of Astronomy, Cornell University, Ithaca, NY 14853, USA}
\email{ayh24@cornell.edu}

\author[0000-0002-5619-4938]{Mansi M. Kasliwal}
\affiliation{Division of Physics, Mathematics, and Astronomy, California Institute of Technology, Pasadena, CA 91125, USA}
\email{mansi@astro.caltech.edu}

\author[0000-0001-6331-112X]{Geoffrey Mo}
\email{gmo@caltech.edu}
\affiliation{Division of Physics, Mathematics, and Astronomy, California Institute of Technology, Pasadena, CA 91125, USA}
\affiliation{The Observatories of the Carnegie Institution for Science, Pasadena, CA 91101, USA}

\author[0000-0003-3173-4691]{Anirudh Salgundi} \affiliation{University of North Carolina at Chapel Hill, 120 E. Cameron Ave., Chapel Hill, NC 27514, USA} \email{anirudhs@unc.edu}

\author[0009-0009-1099-7135]{Kendall I. Sippy} \affil{Department of Astronomy and Astrophysics, The Pennsylvania State University, State College, PA 16802, USA} \email{kxs6464@psu.edu}
 
\author[0000-0003-1546-6615]{J. Sollerman}
\affiliation{The Oskar Klein Centre, Department of Astronomy, Stockholm University, AlbaNova, SE-106 91 Stockholm, Sweden} \email{jesper@astro.su.se}

\author[0000-0001-8018-5348]{Eric C. Bellm}
\affiliation{DIRAC Institute, Department of Astronomy, University of Washington, 3910 15th Avenue NE, Seattle, WA 98195, USA} \email{ecbellm@uw.edu}

\author[0000-0001-9152-6224]{Tracy X. Chen}
\affiliation{IPAC, California Institute of Technology, 1200 E. California Blvd, Pasadena, CA 91125, USA}
\email{xchen@ipac.caltech.edu}

\author[0000-0002-8262-2924]{Michael W. Coughlin} \affiliation{School of Physics and Astronomy, University of Minnesota, Minneapolis, MN 55414}
\email{cough052@umn.edu}

\author[0000-0001-7663-0808]{Michael C. Davis}\affiliation{School of Physics and Astronomy, University of Minnesota, Minneapolis, MN 55414} \email{davi4614@umn.edu}

\author[0000-0002-3137-4633]{Fabio De Colle} \affiliation{Instituto de Ciencias Nucleares, Universidad Nacional Aut{\'o}noma de M{\'e}xico, A. P. 70-543 04510, D. F. Mexico} \email{fabio@nucleares.unam.mx}

\author{Danielle Frostig}
\affil{Center for Astrophysics\:$|$\:Harvard \& Smithsonian, 60 Garden St. Cambridge, MA 02138, USA} \email{danielle.frostig@cfa.harvard.edu}

\author[0000-0003-2624-0056]{Christopher L.~Fryer}
{\affiliation{Center for Nonlinear Studies, Los Alamos National Laboratory, Los Alamos, NM 87545 USA} \email{fryer@lanl.gov}

\author[0000-0002-3168-0139]{Michael J. Graham}
\email{mjg@caltech.edu}
\affiliation{Division of Physics, Mathematics, and Astronomy, California Institute of Technology, Pasadena, CA 91125, USA}

\author[0000-0002-9364-5419]{Xander J. Hall} \affiliation{McWilliams Center for Cosmology and Astrophysics, Department of Physics, Carnegie Mellon University, Pittsburgh, PA 15213, USA} \email{xjh@andrew.cmu.edu}

\author[orcid=0000-0002-0129-806X]{K. -R. Hinds}
\affiliation{Division of Physics, Mathematics, and Astronomy, California Institute of Technology, Pasadena, CA 91125, USA}
\email{khinds@caltech.edu}

\author[0000-0001-9695-8472]{Luca Izzo} \affiliation{INAF, Osservatorio Astronomico di Capodimonte, Salita Moiariello 16, I-80121 Naples, Italy} \affiliation{DARK, Niels Bohr Institute, University of Copenhagen, Jagtvej 128, 2200 Copenhagen, Denmark} \email{luca.izzo@inaf.it}

\author[0000-0002-3934-2644]{Wynn Jacobson-Gal\'an}
\altaffiliation{NASA Hubble Fellow}
\affiliation{Division of Physics, Mathematics, and Astronomy, California Institute of Technology, Pasadena, CA 91125, USA} \email{wynnjg@caltech.edu}

\author[0000-0002-4585-9981]{Nathan~P.~Lourie}
\affiliation{Department of Physics and Kavli Institute for Astrophysics and Space Research, Massachusetts Institute of Technology, 77 Massachusetts
Ave, Cambridge, MA 02139, USA} \email{nlourie@mit.edu}

\author[0000-0003-2611-7269]{Keiichi Maeda}
\affiliation{Department of Astronomy, Kyoto University, Kitashirakawa-Oiwake-cho, Sakyo-ku, Kyoto, 606-8502. Japan} \email{keiichi.maeda@kusastro.kyoto-u.ac.jp}

\author[0000-0003-1227-3738]{Josiah Purdum}
\affiliation{Caltech Optical Observatories, California Institute of Technology, Pasadena, CA 91125, USA} \email{jpurdum@caltech.edu}

\author[0000-0001-7648-4142]{Ben Rusholme}
\affiliation{IPAC, California Institute of Technology, 1200 E. California Blvd, Pasadena, CA 91125, USA} \email{rusholme@ipac.caltech.edu}

\author[0000-0003-2091-622X]{Avinash Singh}
\affiliation{The Oskar Klein Centre, Department of Astronomy, Stockholm University, AlbaNova, SE-106 91 Stockholm, Sweden} \email{avinash21292@gmail.com}

\author[0000-0003-2434-0387]{Robert Stein}
\email{rdstein@umd.edu}
\affiliation{Department of Astronomy, University of Maryland, College Park, MD~20742, USA}
\affiliation{Joint Space-Science Institute, University of Maryland, College Park, MD 20742, USA}
 \affiliation{Astrophysics Science Division, NASA Goddard Space Flight Center, 8800 Greenbelt Rd, Greenbelt, MD 20771, USA}

\begin{abstract}
In March 2026, the Einstein Probe (EP) discovered its most nearby ($z = 0.0343$) Fast X-ray Transient (FXT), EP\,260321a, the first EP FXT to provide a strong match to expectations for X-ray ``shock breakout'' (SBO) emission. Here, we present our multi-wavelength follow-up campaign of EP\,260321a and its broad-line Type Ic (Ic-BL) supernova (SN) counterpart, SN\,2026gzf. We show that our radio follow-up extending over $5.8 - 54.5$~days post-FXT rules out an on-axis jet counterpart of isotropic-equivalent kinetic energy $E_{K} \gtrsim 10^{49}$~erg for circumburst densities $n > 10^{-2}~{\rm cm}^{-3}$ \rr{and assuming microphysical parameters $\epsilon_e = \epsilon_B = 0.1$.} Our radio data also \rr{constrains a median mass-loss rate of $\dot{M} \lesssim 1.2 \times 10^{-5} M_{\odot}~{\rm yr}^{-1}$} for a Wolf-Rayet progenitor. In addition, we derive \rr{SN\,2026gzf's properties, including $^{56}$Ni mass, diffusion timescale, and expansion velocities}, from our \rr{$\sim$nightly-cadence} optical data and compare them with those of optically discovered Type Ic-BL SNe, finding that SN\,2026gzf is well within the 90\% confidence interval across all properties.We further fit SN\,2026gzf's light curve and determine that combined emission from both interaction with CSM and $^{56}$Ni radioactive decay provides the best fit with plausible model parameters. Finally, using the rate of Ic-BL SNe from the ZTF Bright Transient Survey and assuming all Type Ic-BL SNe produce EP\,260321a-like FXTs, we infer an expected rate of EP-detected SBOs of 4.4 - 16 year$^{-1}$. This is inconsistent at the 90\% confidence level with current EP detection rates, potentially indicating that most Type Ic-BL SNe produce less luminous X-ray SBO signals compared to EP\,260321a.
\end{abstract}


\section{Introduction}
The earliest electromagnetic (EM) radiation from core-collapse supernova (SN) explosions is the shock breakout (SBO; e.g., \citealt{Colgate1974, Falk1978, Klein1978, Imshennik1981, Ensman1992, Matzner1999, Nakar2010, Waxman2017}). SBO occurs when the radiation-dominated SN explosion shock crosses the surface of an evolved star, or when the optical depth ($\tau$) ahead of the shock drops below $\tau < c/v_{\rm s}$, where $v_{\rm s}$ is the shock speed. This SBO releases a bright flash that shines at X-ray or UV wavelengths and persists over seconds to $\sim$ hour timescales \citep{Waxman2017}. As the earliest EM signatures of SN explosions, SBOs are direct probes of the very early stages of the core-collapse process, providing critical insight into the progenitor star mass and composition, explosion time, and rate of mass loss prior to explosion.

Despite their potential as probes of the pre-explosion environment, SBOs are rarely detected as they require observations of the SN at nearly exactly the time of explosion. Though wide-field optical surveys have dramatically increased the rate of SN discoveries (e.g., \citealt{Law2009, Bellm+19, ATLAS_tonry+18, Cochanek2017}); SBOs are expected to peak at temperatures $\sim 0.1$ keV \citep{Nakar2010} and are not expected produce optical emission during the breakout phase. Therefore, wide-field UV and soft X-ray detectors provide the only avenue to discover SBOs from a systematic perspective. 

Prior to 2024, a number of fast X-ray transients (FXTs) had been discovered by soft X-ray satellites such as HETE-2,  Beppo-Sax, and MAXI  \citep{ Heise+01,Soderberg+05,Sakamoto+05, MAXI} or in archival searches \citep{Jonker2013,Glennie2015,Bauer2017,AlpLarsson20,Novara+20,2022ApJ...927..211L,QuirolaVasquez+23,Vasquez2025,Brightman+26}. However, the vast majority lacked counterparts or redshift information, precluding definitive constraints on their origins or connection to SBO. 

Thus far, the only bona fide X-ray SBO event was discovered serendipitously by the \textit{Neil Gehrels Swift Observatory}'s (\textit{Swift}) X-ray Telescope (XRT; \citealt{Swift-XRT}), which operates from 0.3 -- 10 keV. During a scheduled XRT observation of galaxy NGC 2770, a bright X-ray flash (XRF\,080109; XRFs belong to the broader class of FXTs) was detected in one of the spiral arms in the galaxy, lasting $\sim 400$ s \citep{Soderberg+08}. UV observations by \textit{Swift}'s Ultraviolet Optical Telescope (UVOT; \citealt{Swift-UVOT}) 1.4 hours after the X-ray outburst revealed a brightening UV and optical source, and later optical spectroscopic observations confirmed the associated optical transient was SN 2008D \citep{Modjaz+09, Malesani2009, Maund2009, Mazzali2008, Chevalier2008, Soderberg+08}. SN 2008D initially showed broad absorption features in its optical spectrum with no H or He lines \citep{Modjaz+09}, reminiscent of broad-lined Type Ic SNe (Type Ic-BL; \citealt{Filippenko1997, Taddia2019, Srinivasaragavan+24}). However, SN\,2008D evolved to show strong He absorption features, leading to the eventual classification of a Type Ib SN \citep{Soderberg+08,Modjaz+09, Maund2009}. 

The discovery of XRF\,080109/SN 2008D was ground-breaking, but its serendipitous nature and the absence of additional SBOs made it clear that a wide-field, sensitive UV/soft X-ray satellite was necessary to accrue a sample of SBOs. The Einstein Probe (EP; \citealt{Yuan+22,Yuan2025}), or Tianguan mission, which began operations in 2024, is well-suited to regular detections and prompt alerts of new FXTs. EP's Wide-field X-ray Telescope (WXT), which has an instantaneous field of view of 3850 deg$^2$, operates in the soft X-rays from 0.5 to 4 keV.

In just two years of operations, EP has detected $\gtrsim 100$ FXTs. Until recently, only five had been associated with SN Ic-BL counterparts (EP\,240414a, 240801a, 250108a, 250304a, 250827b; \citealt{Sun+24,vanDalen+24, EylesFerris25,Hamidani+25,Li+25, Rastinejad+25, Zheng+25, Srinivasaragavan+25, Srinivasaragavan25b,SCG+25,Cotter:2026yus,vanHoof+26}). The origins of the X-ray emission for these events remain ambiguous (e.g., \citealt{EylesFerris25,Srinivasaragavan+25,Zheng+25}). However, they are generally more luminous and spectrally harder compared to thermal SBO models, disfavoring a ``classical'' (non-thermal, non-relativistic) SBO origin. Instead, these FXTs may be the product of relativistic jets, in line with observations of several FXTs with associated with long gamma-ray bursts (GRBs; \citealt{Liu+25, Ricci+25, Yin+24}).

In this Letter, we present our multi-wavelength follow-up campaign and analysis of EP\,260321a, the first EP FXT with properties consistent with expectations for SBO \citep{GCNanalysis}. In line with previously reported observations (e.g., \citealt{2026GCN.44082....1T,2026GCN.44070....1L}), we show that EP\,260321a originated in the galaxy SDSS J095942.88+002506.2 at $z = 0.0343$ (156~Mpc) and was accompanied by SN\,2026gzf, rendering it the most nearby EP FXT with an associated SN discovered to date. In Section~\ref{sec:obs} we present our new optical and radio observations of EP\,260321a's counterpart. In Section~\ref{sec:analysis} we present analytic evidence that EP\,260321a was the product of SBO, and infer properties of the circumstellar material (CSM) surrounding the progenitor star. In Section~\ref{sec:sn2026gzf} we infer the properties of SN\,2026gzf and contrast these with other stripped-envelope SNe (SESNe; Types Ib, Ic, Ic-BL), showing that SN\,2026gzf is comparable to optically discovered Type Ic-BL SNe. In Section~\ref{sec:host} we perform an analysis of EP\,260321a's host galaxy, SDSS J095942.88+002506.2. Finally, in Section~\ref{sec:discussion} we review the combined picture of CSM surrounding the progenitor star from our multi-wavelength probes and estimate the rate of EP-detected SBO signals. Throughout this work, we present all magnitudes in AB units and corrected for Galactic extinction of E(B-V) = 0.02 (unless otherwise specified; \citealt{SchlaflyFinkbeiner11}), and assume a Planck cosmology \citep{Planck20}.

\section{Observations}
\label{sec:obs}

\subsection{X-ray}
\label{sec:X-ray}
EP's WXT triggered on EP\,260321a on UT 2026-03-21 12:23:07 (hereafter $t_0$; \citealt{GCNDiscovery}). The event lasted for $\sim$432 seconds. \rr{Observations were} subsequently interrupted by EP's Follow-up X-ray Telesocpe (FXT) automated follow-up observation of the field \citep{GCNanalysis}\footnote{The \textit{Fermi} Space Telescope was in the South Atlantic Anomaly at the time of the WXT trigger, and thus does not provide an upper limit on $\gamma$-ray emission from the source. \textit{Swift}'s Burst Alert Telescope was not operating at the time of the trigger.}. The average unabsorbed flux of the WXT observation in the 0.5 -- 4 keV band was $8.0 ^{+2.2}_{-1.9} \times 10^{-11} \, \rm{erg \, s^{-1} cm^{-2}}$. At the redshift $z = 0.0343$, the average flux corresponds to a peak luminosity of $\sim 2.2 \times 10^{44} \, \rm{erg \, s^{-1}}$. The WXT detection can be fit with a blackbody spectrum, with a peak energy $E_p = 0.164^{+0.04}_{-0.029}$ keV \citep{GCNanalysis}.

\begin{figure}
    \centering
    \includegraphics[width=\linewidth]{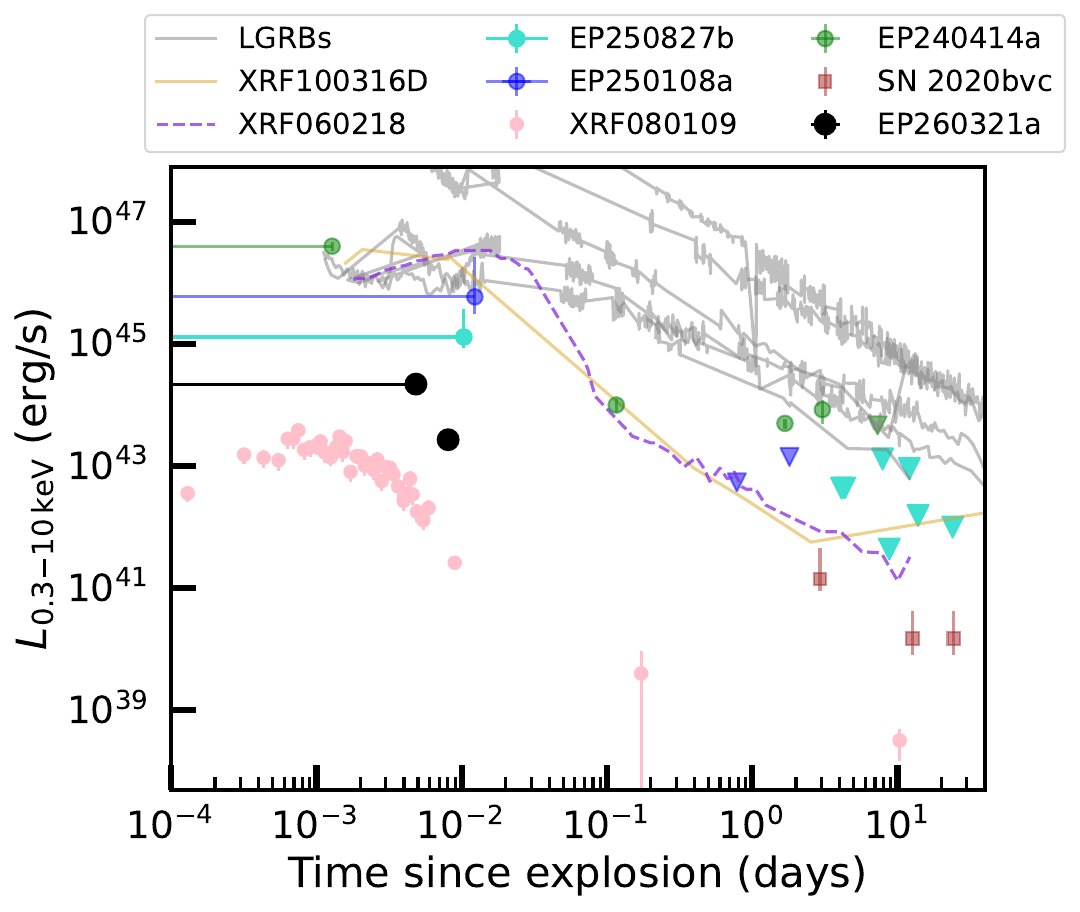}
    \caption{Comparison of the X-ray properties of EP\,260321a from the GCNs \citep{GCNanalysis} against the X-ray light curves of other SESNe \rr{\citep{Evans+07,Evans+09,Ho+20,Sun+24,Srinivasaragavan+25,Srinivasaragavan25b}}. With the exception of SBO XRF\,080109, EP\,260321a is the least luminous FXT, peaking several orders of magnitude below FXTs associated with on-axis relativistic jets.}
    \label{fig:xray_gcn}
\end{figure}

\begin{figure*}[!t]
\centering
\includegraphics[width=0.42\textwidth]{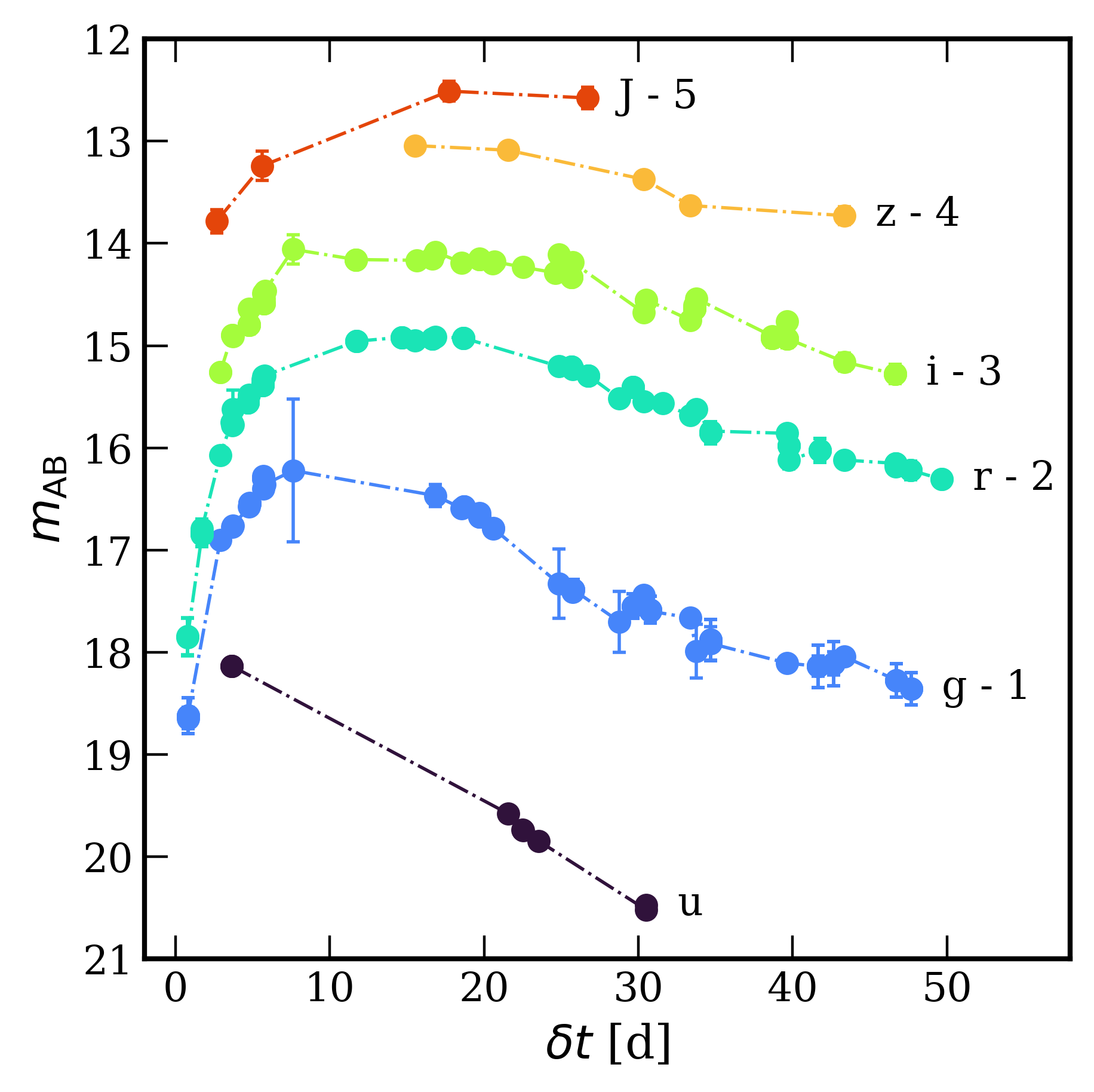}
\includegraphics[width=.57\textwidth]{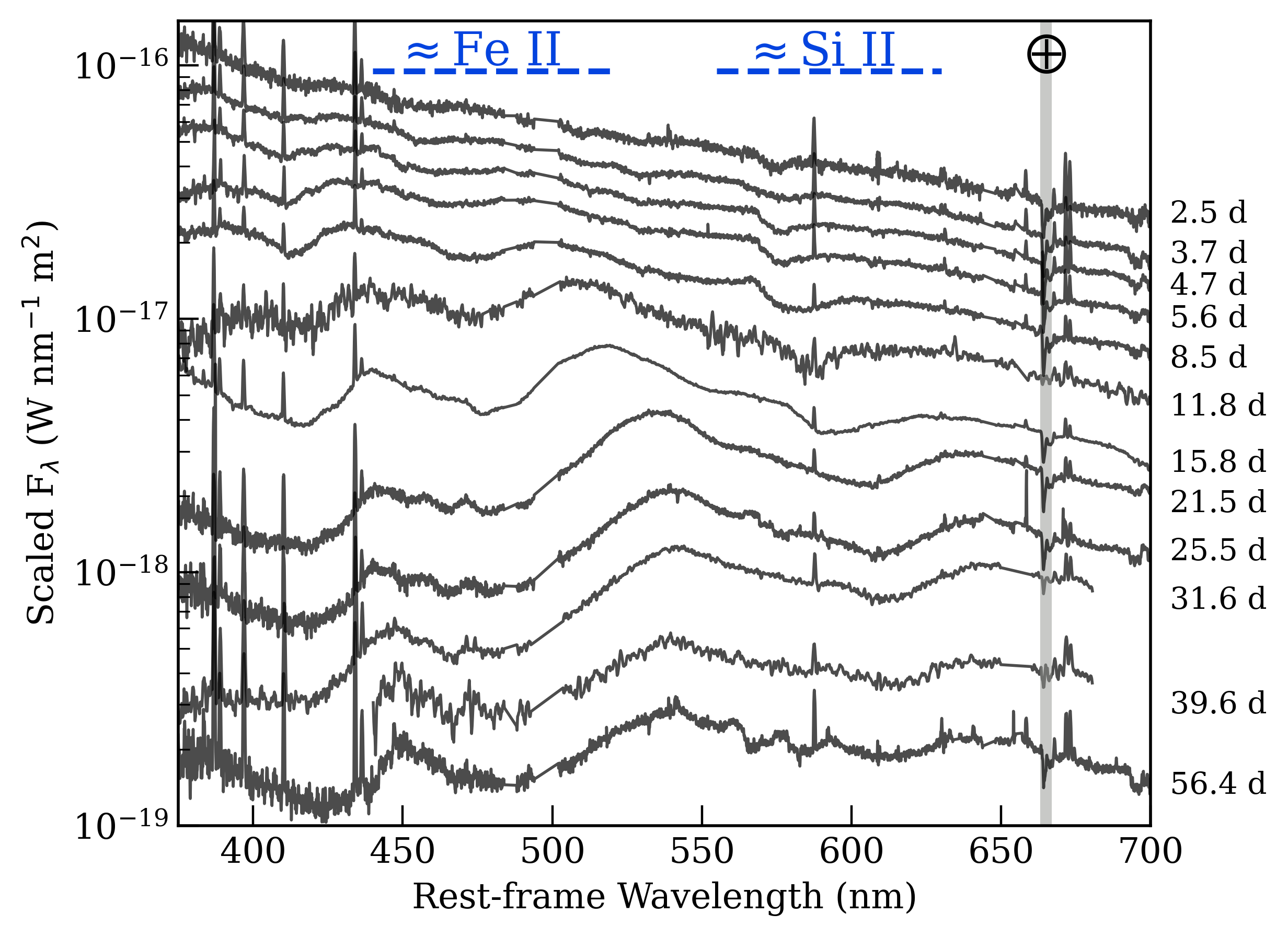}
\caption{Optical observations of the Type Ic-BL counterpart to EP\,260321a, SN\,2026gzf. \textit{Left:} Our optical and near-IR light curve of SN\,2026gzf. \textit{Right:} Sequence of GMOS, SOAR and NGPS spectra of SN\,2026gzf. We mask the regions of the emission lines of H$\alpha$, H$\beta$, [O II] for visibility. Spectra are not corrected for Galactic extinction, which is weak along the line-of-sight (E(B-V) = 0.02; \citealt{SchlaflyFinkbeiner11}).}
\label{fig:opt_lc_spec}
\end{figure*}

FXT then observed EP\,260321a at $\delta t =$12 minutes (where $\delta t$ is observed-frame time since $t_0$) and found an uncatalogued fading source within the WXT error circle at  $\alpha$ (J2000)=149.9287\degree and $\delta$ (J2000) = 0.4177\degree (10 \arcsec uncertainty radius; \citealt{GCNanalysis}). The average, unabsorbed flux in the 0.5 -- 10 keV band was
($9.8 \pm 0.3) \times 10^{-12} \, \rm{erg \, s^{-1} cm^{-2}}$ \citep{GCNanalysis}. The FXT spectrum was well-fit with a blackbody model with a peak energy $E_p = 0.121^{+0.03}_{-0.03}$ keV \citep{GCNanalysis}. \citet{GCNanalysis} reported that its soft spectrum, rapid decay, and high luminosity rendered EP\,260321a a promising SBO candidate.

In Figure~\ref{fig:xray_gcn} we contrast the X-ray light curve of EP\,260321a against previous EP FXTs, \textit{Swift} XRFs, and the X-ray light curves of LGRBs and Type Ic-BL SN\,2020bvc \citep{Izzo+20,Ho+20}. With the exception of SBO XRF\,080109, EP\,260321a is the least luminous FXT, peaking nearly an order of magnitude below those of all previous FXT/XRF/GRB events. Overall, we observe a wide span in X-ray luminosity, nearly ten orders of magnitude at $<1$~day post-explosion, across a relatively small sample of events.

\subsection{Photometry}
\label{sec:phot}

Within the FXT localization, optical follow-up quickly discovered a blue, variable point source coincident with a star-forming galaxy at $z = 0.0343$ ($d_L~= 156$~Mpc; \citealt{2026GCN.44082....1T,2026GCN.44070....1L}). The Zwicky Transient Facility (ZTF; \citealt{Bellm+19,Graham+19,Masci+19,Dekany+20}) detected the optical counterpart to EP\,260321a, ZTF26aaonmha, in its regular surveying mode at $\delta t = 0.78$~days. In Sections~\ref{sec:spec} we present evidence that this source is a SN counterpart to EP\,260321a, SN\,2026gzf, which resides in a blue ``knot'' within SDSS J095942.88+002506.2. The source was observed to be coincident with an existing compact, blue ``knot'' observed in Legacy Survey archival imaging \citep{Dey+19}, which we discuss further in Section~\ref{sec:host}. 

We obtained optical photometry of SN\,2026gzf in the $ugriz$-bands with ZTF, the Spectral Energy Distribution machine (SEDm; \citealt{Blagorodnova+18,Rigault+19}), and IO:O \citep{LT_IOO} mounted on the Liverpool Telescope (Program ID: XJL24B16; PI: Bochenek) through the Fritz.science instance of SkyPortal \citep{vanderWalt+19,Coughlin+23} over $\delta t = 0.78-49.6$~days. We further obtained $ugriz$ photometry of SN\,2026gzf observed by the Legacy Survey of Space and Time (LSST) conducted by the Vera Rubin Observatory (Rubin; \citealt{Ive19,Rubin_alerts,JegouduLaz+25}). Both ZTF and LSST observed variability ($\gtrsim$3-5$\sigma$ detections) at the location of SN\,2026gzf several years to $\sim 1$ month prior to EP\,260321a, potentially indicative of variability in the progenitor star. However, upon further examination we are unable to confirm any of these pre-FXT detections as real (see Appendix~\ref{sec:phot_red}). We obtained additional $J$-band photometry of EP\,260321a with WINTER \citep{Frostig+22}. We describe our photometric data reduction in Appendix Section~\ref{sec:phot_red}, show our light curves in Figure~\ref{fig:opt_lc_spec}, and report our photometry in Appendix Table~\ref{tab:phot}.

We fit a low-order polynomial to our $r$-band light curve of SN\,2026gzf, finding a peak \rr{absolute magnitude} of $M_{r} = -19.1$ at $\delta t =$ 14.2~days. We compare the properties of SN\,2026gzf to the wider sample of SESNe in Section~\ref{sec:comp}. We determine the position of SN\,2026gzf using acquisition imaging taken for the early GMOS spectra (Section~\ref{sec:spec}). We stack the 60~s acquisition images taken at $\delta t = 3.7, 4.7, 5.6, 8.5$ and 11.8~days with the pipeline Data Reduction for Astronomy from Gemini Observatory North and South (``DRAGONS''; \citealt{DRAGONS19}), apply astrometry using astrometry.net \citep{lang10}, and determine SN\,2026gzf's position using Source Extractor \citep{Bertin1996}. The final optical position is $\alpha$=09$^{\rm h}$59$^{\rm m}$42.8751$^{\rm s}$, $\delta=$00\arcdeg25\arcmin06.454\arcsec, with a 3$\sigma$ uncertainty radii of 0.375\arcsec.

\subsection{Spectroscopy}
\label{sec:spec}

We obtained optical spectroscopy of SN\,2026gzf over $\delta t = 2.5 - 56.4$~days with the Gemini Multi-Object Spectrographs (GMOS; \citealt{GMOS04}) mounted on the Gemini-North and South telescopes (Programs GN-2026A-Q-208, GS-2026A-Q-119, GS-2026A-Q-203; PIs: Rastinejad, Srinivasaragavan), the Next Generation Palomar Spectrograph (NGPS; \citealt{NGPS}) mounted on the Hale telescope (PI: Fremling), and the Goodman spectrograph \citep{Clemens2004} mounted on SOAR (Program SOAR2026A-018; PI: Andreoni). We further obtained a near-IR spectrum of SN\,2026gzf at $\delta t = 30$~days with the Folded-port Infrared Echellette (FIRE; \citealt{Simcoe+13}) spectrograph on Magellan (PI: Anand). We describe our spectroscopic set-up and data reduction in the Appendix Section~\ref{sec:spec_red} and present SEDm spectra, which are not used in our analysis as they provide similar temporal coverage and coarser spectral information, in Appendix Figure~\ref{sedmspec}. 

In Figure~\ref{fig:opt_lc_spec} we show our optical spectral series, smoothed using a Savisky-Golay filter. In all spectra, we clearly detect strong emission lines from the underlying host galaxy, including H$\alpha$, H$\beta$, H$\delta$, H$\gamma$, [O III], and [S II], confirming the event distance at $z = 0.0343$. Our spectra do not show significant reddening nor Na I D absorption, indicating minimal dust along the line-of-sight from the host galaxy. We estimate the equivalent width (EW) of the Na I D line to be EW$\lesssim0.167$ from the GMOS spectrum obtained on 2026-04-12, following \citep{Das+23}. Since Na I D absorption is not consistently detected across all spectroscopic epochs, we treat the above estimate as an upper limit. Assuming the relation between host galaxy extinction and Na I D EW, $A_\mathrm{v}[\mathrm{mag}]=0.78(\pm0.15) \times \mathrm{EW_{Na\,I\,D}}[\AA]$ \cite{Stritzinger+18}, we calculate a host extinction upper limit of $A_\mathrm{v} < 0.13$\,mag. 

The spectra at early times ($\delta t \sim 3.7 - 8.5$~days) are remarkably blue, and show weak, broad absorption features around $\approx$ 3800 - 4900~\AA\, and $\approx$ 5200 - 6000~\AA, which we ascribe to Fe II $\lambda 5169$, and Si II $\lambda 6355$. These broad absorption features become more prominent at $\delta t \gtrsim 12$~days. Across all epochs, our spectra do not show any obvious absorption at the expected locations of H or He lines, in contrast to SN\,2008D \citep{Soderberg+08,Modjaz+09}. We classify our GMOS spectrum at $\delta t = 5.7$~days using SNID \citep{Blondin+07} and find a best-match to the Type Ic-BL SN\,2006aj at 3~days post-peak. SN\,2006aj was associated with the GRB/XRF\,060218, whose peak X-ray luminosity was a factor of $\sim$100 brighter than EP\,260321a (Figure~\ref{fig:xray_gcn}; \citealt{Campana+06,Soderberg+06,Modjaz+06}). We therefore conclude that SN\,2026gzf is a Type Ic-BL SN.

\subsection{Radio}
\label{sec:radio}

We observed the field of SN\,2026gzf with the VLA in its A configuration over five epochs. The first and fifth epochs were obtained on March 27 ($\delta t = 5.7$~days; 6, 10, 15 and 22 GHz) and May 15 ($\delta t = 54.5$~days; 6 and 10 GHz) under program 26A-385 (PI: Leung). We obtained three additional epochs at 6 GHz under program 25B-282 (PI: O'Dwyer) on April 4 ($\delta t = $ 12.8 days), April 13 ($\delta t = $ 22.0 days) and May 11 ($\delta t = $ 50.7 days). For all data, we use J1007-0207 as a phase calibrator and 3C286 for flux density and bandpass calibration. We use J1024-0052 for pointing calibration in our \rr{higher-frequency 15 and 22\,GHz observations} under program 26A-385 \rr{because our phase calibrator was not bright enough for this purpose}.

\begin{figure}
    \centering
    \includegraphics[width=\linewidth]{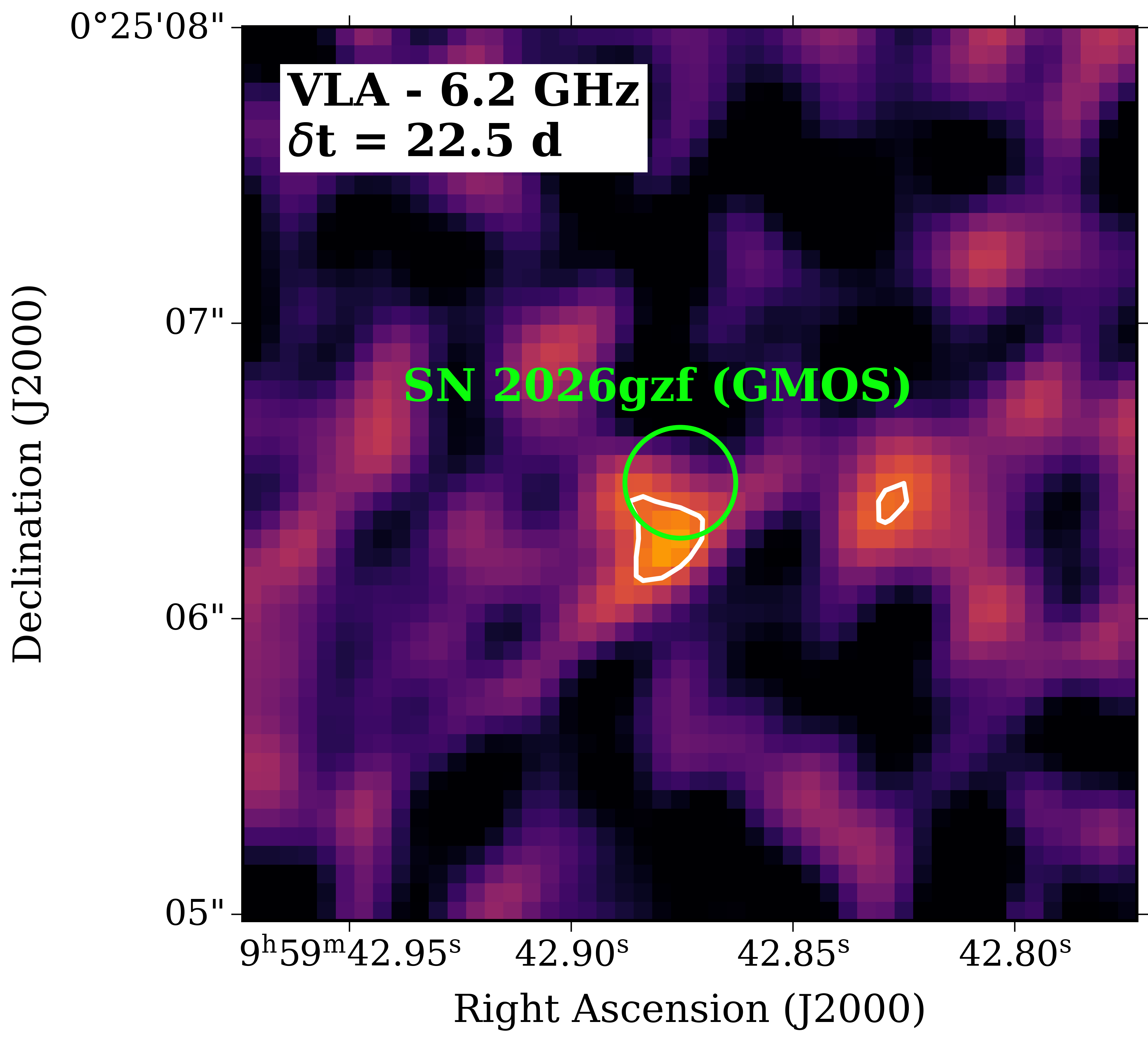}
    \caption{VLA 6.2~GHz image of the field of SN\,2026gzf. White contours denote 3$\sigma$ radio sources in the vicinity of SN\,2026gzf, while the green ellipse shows SN\,2026gzf's position measured from our GMOS imaging. The centers of the radio source and SN\,2026gzf are offset 0.197\arcsec. The slight offset and absence of significant variation in the radio flux across epochs is consistent with underlying star formation rather than a radio counterpart to EP\,260321a.}
    \label{fig:radio_image}
\end{figure}

\rr{We use the NRAO pipeline-calibrated images  \citep{2020ASPC..527..571K} in \texttt{CASA} \citep{2022PASP..134k4501C}, comparing both  standard calibrated continuum science-target data (\texttt{regcal}) and the corresponding products with self-calibration applied (\texttt{selfcal}) to calibrate and image the data, Flux densities were comparable within 1$\sigma$ thus (\texttt{selfcal}) value are reported in Table \ref{tab:vla_observations}}. Images were cleaned down to the $5\sigma$ level and automated self-calibration on the science targets was carried out using auto-masking. After automated calibration and imaging, all data were also manually inspected for the presence of potential radio frequency interference and problematic antennae with no outstanding issues found. 

A faint radio counterpart to SN\,2026gzf is detected at $>3\sigma$ in nearly all images at 6~GHz (Figure~\ref{fig:radio_image}). For the radio epochs at $\sim 13-51$ days, \texttt{CASA imfit} measurements each returned a point like source component that may be as large as $0.36\,\text{arcsec} \times 2\,\text{arcsec}$ with an integrated flux of $24.4 \pm7.3 \mu$Jy. The position is slightly offset (0.197\,\arcsec) from our GMOS position (Figure~\ref{fig:radio_image}). Given this offset and that the measured radio flux density at 6.2~GHz days remains unchanged within the errors across all epochs, we favor emission from star formation in the local environment as its source. 

We employ measurements of this emission ($F_{\nu}$ in Table~\ref{tab:vla_observations}) as upper limits in our analysis and estimate the underlying star formation rate (SFR) from our radio observations in Section~\ref{sec:radio_sfr}. We measure radio flux densities in the VLA images using the \texttt{imstat} tool in \texttt{CASA}. For each VLA epoch, we report the maximum flux density found within a circular region centered on $\alpha$=09$^{\rm h}$59$^{\rm m}$42.872$^{\rm s}$, $\delta=$00\arcdeg25\arcmin06.38\arcsec, with a radius equal to the nominal VLA synthesized beam at the configuration and frequency setup of each observation (0.33\arcsec\ for A-config C-band). The root-mean-square (RMS) of the noise is estimated from each image using a circular region of radius $\approx 10\times$ the nominal FWHM of the synthesized beam centered on the position of SN\,2026gzf. We list all VLA observations in Table \ref{tab:vla_observations}. We incorporate additional radio observations from the GCNs in our analysis \citep{ep260321a_uGMRT,ep260321a_ATCA}.

In Figure~\ref{fig:radio_comp} we plot our 6~GHz upper limits against the radio light curves of previous GRBs, FXTs, XRF\,080109, and Type Ic-BL SN\,2020bvc \citep{Kulkarni+98,Waxman+98,Soderberg+04,Soderberg+06,Soderberg+08,Bietenholz+09,Ho+20,Bright+24_EP240414,Srinivasaragavan+25}. Our upper limits probe lower luminosities compared to all events shown. We further use these upper limits to constrain synchrotron emission from a relativistic on-axis jet and fast-moving SN ejecta in Section~\ref{sec:analysis}.

\begin{table}[ht]
\centering
\begin{tabular}{ccccc}
\toprule
$t_{\rm mid}$ & $\delta t$ & $\nu$ & $F_{\nu}$  & Image RMS \\
(MJD) & (d) & (GHz) & ($\mu$Jy) & ($\mu$Jy) \\
\midrule
61126.25 & 5.7 & 22.0 & 17 & 5.8 \\
61126.27 & 5.7 & 14.9 & 8.0 & 7.2 \\
61126.29 & 5.8 & 9.7 & 19 & 7.7 \\
61126.31 & 5.8 & 6.2 & 19 & 5.8 \\
61134.17 & 12.8 & 6.2 & 19 & 4.7 \\
61143.06 & 22.5 & 6.2 & 20 & 4.9 \\
61172.06 & 50.7 & 6.2 & 23 & 4.7 \\
61175.05 & 54.5 & 9.8 & 21 & 8.5 \\
61175.07 & 54.5 & 6.2 & 15 & 7.3 \\
\bottomrule
\end{tabular}
\caption{VLA observations covering the position of EP\,260321a/SN\,2026gzf. As the measured 6~GHz flux (F$_{\nu}$) is consistent across all epochs, we assume the radio emission is dominated by underlying star formation. We use F$_{\nu}$ as an upper limit in our modeling.}
\label{tab:vla_observations}
\end{table}

\begin{figure}
\centering
\includegraphics[width=0.49\textwidth]{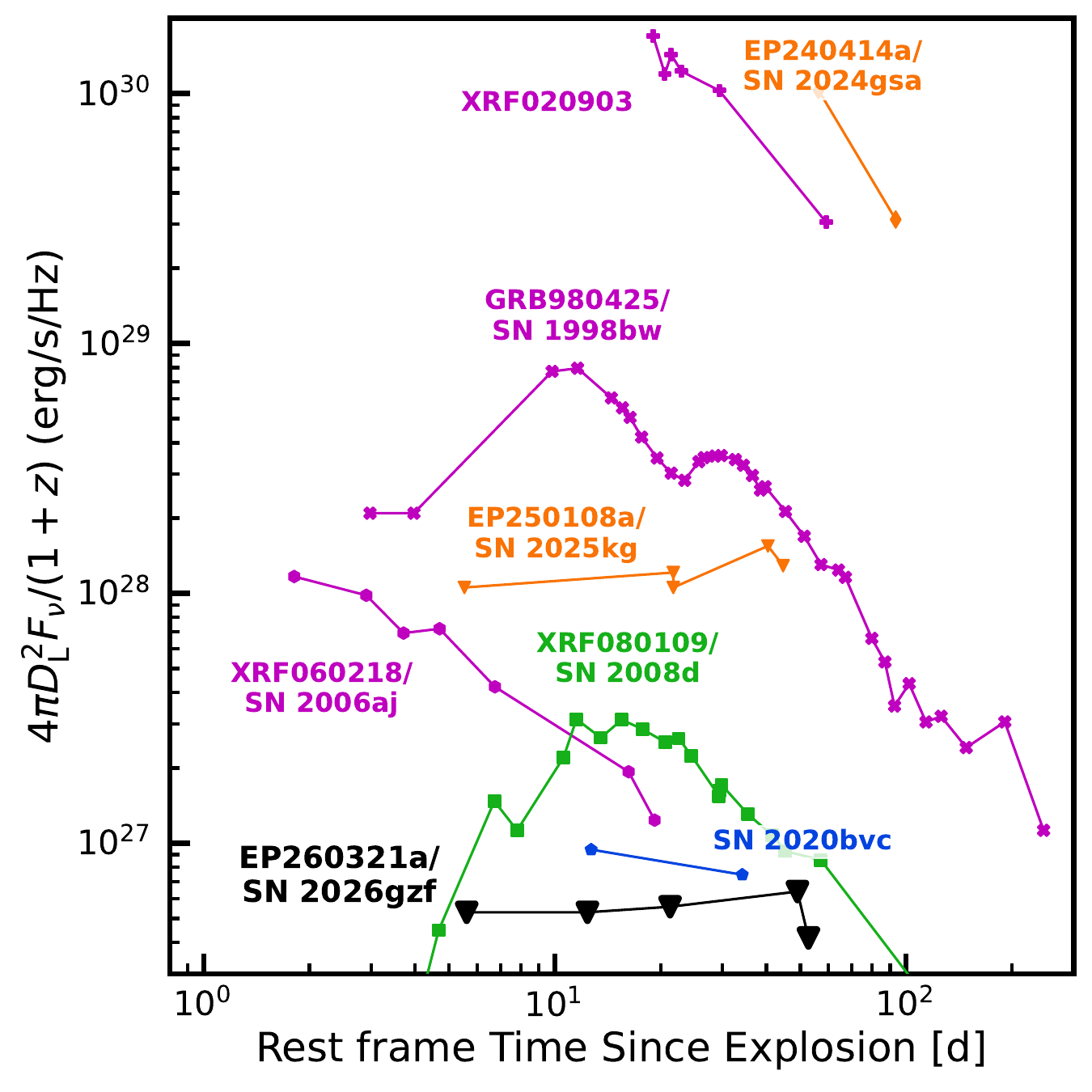}
\caption{Comparison of radio upper limits of EP\,260321a/SN\,2026gzf with other GRB, FXT and Type Ic-BL SNe \citep{Kulkarni+98,Waxman+98,Soderberg+04,Soderberg+06,Soderberg+08,Bietenholz+09,Ho+20,Bright+24_EP240414,Srinivasaragavan+25}. Our upper limits probe lower luminosities compared to all previous events shown.}
\label{fig:radio_comp}
\end{figure}

\section{Analysis of X-ray and Radio Components}
\label{sec:analysis}

\subsection{Radio Constraints on On-Axis Jet Energetics}

The detection of GRBs coincident with some EP FXTs demonstrates that on-axis relativistic jets are a source of some FXTs (e.g., \citealt{Yin+24,GCN.37071,Liu+25}). We are thus motivated to investigate whether EP\,260321a is the product of an on-axis jet. A signature of such jets is broadband synchrotron afterglow emission, observations of which can constrain the jet's isotropic-equivalent energy ($E_{\rm K, iso}$) and circumburst density ($n$; e.g., \citealt{GranotSari02}). We use the afterglow modeling software \texttt{vegasafterglow} \citep{Wang+26_VegasAfterglow} implemented in \textsc{Redback} to constrain the range of on-axis jet energies allowed by our deep radio upper limits (Section~\ref{sec:radio}). 

We model the radio data as 3$\sigma$ upper limits using the \textsc{Redback} Gaussian upper limit likelihood. We assume an on-axis top-hat jet expanding into a constant-density ISM. We fix the viewing angle to $\theta_{\rm obs}=0$, set the wind-like density parameter to a negligible value, and fix the microphysical parameters to $\epsilon_e=\epsilon_B=0.1$. We adopt broad priors on the isotropic-equivalent kinetic energy and density, $\log_{10}(E_{\rm K,iso}/{\rm erg})=48$--$54$ and $\log_{10}(n/{\rm cm^{-3}})=-5$--$-2$, and allow the jet opening angle, initial Lorentz factor, and electron energy distribution index to vary over
$\theta_c=0.05$--$0.25$ rad, $\Gamma_0=10$--$1000$, and $p=2.5$--$3.0$. We emphasize that as the radio data are all non-detections, the posterior should be interpreted as the region of parameter space that remains allowed by the limits under these assumptions, rather than as any strong constraint on the properties of a potential afterglow.

The radio upper limits strongly disfavor a normal energetic on-axis GRB-like afterglow. The allowed on-axis top-hat posterior has $\log_{10}(E_{\rm K,iso}/{\rm erg})=48.4$ with a 90\% credible interval of $48.0$--$49.1$, corresponding to a 95th-percentile upper bound of
$E_{\rm K,iso}\simeq1.2\times10^{49}$~erg. We show the $\log_{10}(E_{\rm K,iso}/{\rm erg})-\log_{10}(n/{\rm cm^{-3}})$ parameter space allowed by our limits in Figure~\ref{fig:radio_constraints} and the full corner plot in \rr{Appendix Figure~\ref{fig:corner_jet}}. Restricting to densities $n>10^{-2}~{\rm cm^{-3}}$, the 95th-percentile upper bound is
$E_{\rm K,iso}\simeq7.8\times10^{48}$~erg. Higher energies are only possible for very low-density environments, and densities $n\lesssim10^{-2}~{\rm cm^{-3}}$ are uncommon for long GRBs \citep{PanaitescuKumar02,KangasFruchter21,Schroeder+25}, which are likely to trace broadly similar massive-star environments to Type Ic-BL SN\,2026gzf. These limits therefore disfavor an ordinary on-axis GRB jet as the origin of EP\,260321a. 

Our radio observations span a limited time-window $\delta t=5.7$--$54.5$~d after explosion and are therefore less constraining for off-axis jets, structured jets, or jets in extremely low-density environments. We find that an off-axis top-hat jet with standard microphysics with $\theta_{\rm obs}/\theta_c\simeq7.7$ and a broad allowed $E_{\rm K,iso}$ range extending to normal GRB-like energies is allowed by our upper limits. A Gaussian structured jet gives a similar conclusion, with $\theta_{\rm obs}/\theta_c\simeq7.0$ and broad allowed energies. An on-axis jet can also be hidden if the microphysical parameters are allowed to be small, with representative posterior values $\log_{10}\epsilon_e\simeq-2.0$ and $\log_{10}\epsilon_B\simeq-4.0$. By contrast, an on-axis GRB or low-luminosity GRB \rr{($E_{\rm K,iso} \gtrsim 10^{49}$~erg; \citealt{Cano+17_review})} with standard microphysics survives only in an extremely low-density environment, $\log_{10}(n/{\rm cm^{-3}})\simeq-7.7$, which we regard as physically implausible given SN\,2026gzf's environment (Section~\ref{sec:host}). \rr{We further find that allowing $\epsilon_B$ to vary freely over $-4.0\leq\log_{10}\epsilon_B\leq-1.0$ only weakens the 95th-percentile upper limit to $E_{\rm K,iso} \leq 5 \times 10^{49}$~erg for $n\lesssim10^{-2}~{\rm cm^{-3}}$ (Appendix Figure~\ref{fig:corner_jet_epsBfree}), still below a typical on-axis GRB afterglow \citep{Cano+17_review}.} Thus, the radio data rule out an
ordinary on-axis GRB-like afterglow under standard assumptions (Figure~\ref{fig:radio_constraints}), but they do not exclude an off-axis jet, a structured weak jet, or a jet with unusually low microphysical efficiencies. Late-time radio observations \rr{will probe an off-axis jet, which may be distinguished by its light curve shape.}

\begin{figure*}
    \centering
    \includegraphics[width=.7\textwidth]{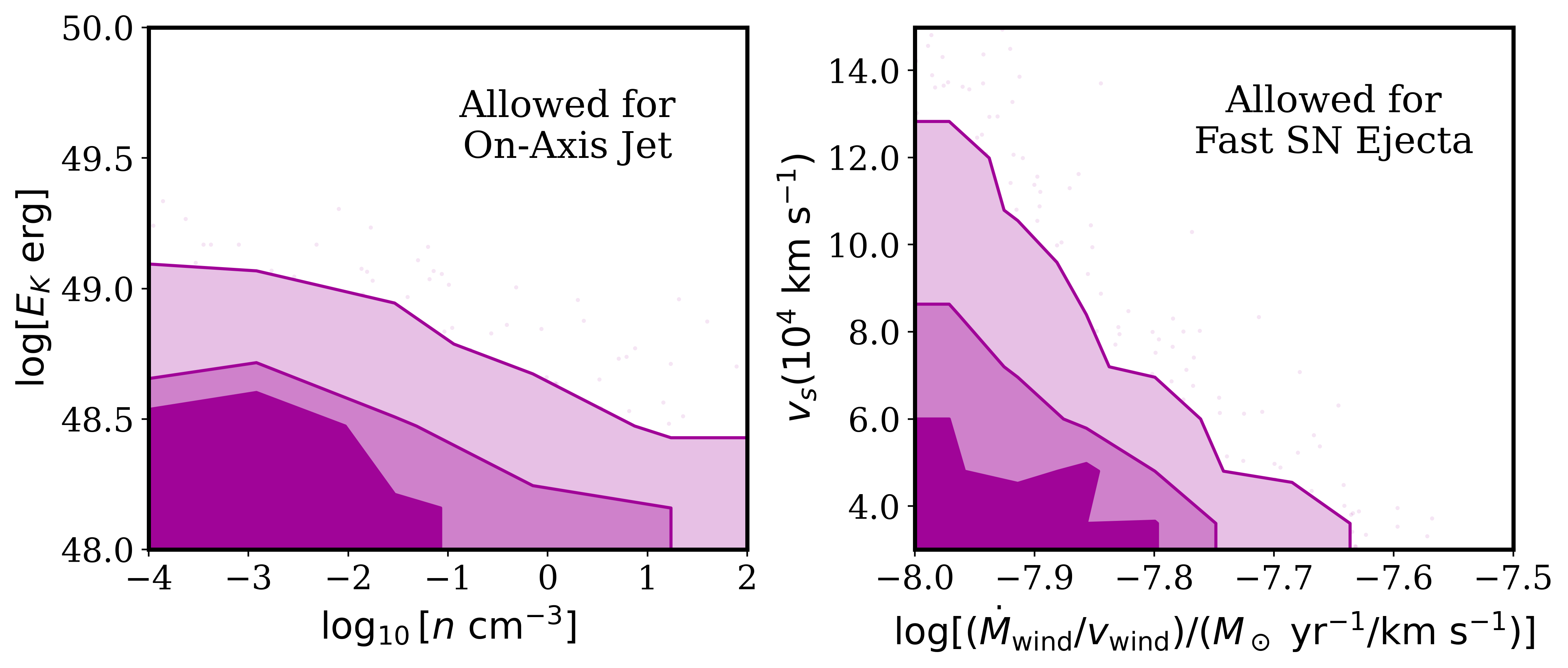}
    \caption{Constraints on models for synchrotron emission from an on-axis top-hat jet expanding into an ISM-like environment (left) and fast SN ejecta interacting with a wind-like CSM (right) from our radio upper limits. Contours show the 50\%, 68\% and 90\% probability regions. The radio limits constrain an on-axis jet to
    $E_{\rm K,iso}\lesssim10^{49}$~erg for typical massive star environments ($n > 10^{-2}~{\rm cm^{-3}}$) and imply low circumstellar mass loading, with fast-ejecta models favoring
    $\log_{10}[(\dot{M}/v_w)/(M_\odot~{\rm yr^{-1}}/{\rm km~s^{-1}})]
    \simeq -7.8$}
    \label{fig:radio_constraints}
\end{figure*}

\subsection{Analytic Evidence for X-ray SBO}
\label{sec:analytic_constraints}

We next investigate whether SBO, which was strongly favored to explain the FXT preceding SN\,2008D \citep{Soderberg+08,Mazzali2008,Modjaz+09}, is a plausible explanation for EP\,260321a. We use the treatment presented in \citet{Haynie2021} to probe whether emission due to sub-relativistic SBO of ejecta in a CSM environment can adequately describe the WXT detection (see \S \ref{sec:X-ray}), assuming that diffusion is the dominant rate-limiting process. The SBO's rise time can be represented as  
\begin{equation}
    t_r = \frac{R_{\rm{e}}^2}{R_d v_t}
\end{equation}
where $R_{\rm{e}}$ is the radius of the CSM and $v_t$ is the characteristic velocity of the shock. \rr{$t_r$ is defined by how quickly photons can diffuse out of the extended envelope from the depth of the SBO, $R_d$}. $R_d$ is defined as
\begin{equation}
    R_d = \frac{\kappa_{C} D v_t}{c}
\end{equation}
where $\kappa_C$ is the opacity of the CSM, and $D$ is the mass loading factor, 
\begin{equation}
   D = \frac{\dot M}{4\pi v_t}.
\end{equation}
Since $M_{\rm{e}}$, the mass of CSM, $\sim 4\pi R_{\rm{e}} D$, we integrate the expression for $D$ with respect to mass and substitute into the expression for $R_d$ to get 
\begin{equation}
    R_d = \frac{\kappa_C v_t}{c} \frac{M_{\rm{e}}}{4\pi R_{\rm{e}}}.
\end{equation}
After substituting this expression into $t_r$, we find 
\begin{equation}
\label{eqtime}
    t_r = \frac{4 \pi R_{\rm{e}}^3 c}{\kappa_C M_{\rm{e}} v_t^2}.
\end{equation}

Since we are in the diffusion-dominated regime, the luminosity of the SBO is 
\begin{equation}
    L_{\rm{SBO}} = \frac{E_{\rm{SBO}}}{t_r} = \frac{4\pi c R_e v_t}{\kappa}\approx 4\pi c v_t^3,
\label{eqlum}
\end{equation}
where $E_{\rm{SBO}}$ is the SBO's energy \rr{(see \citet{Haynie2021} for a full derivation)}.

If the X-ray prompt emission is generated through the SBO, the SBO's rise time should be similar to the rest-frame timescale of the prompt emission. We only have a lower limit on this timescale of $\sim 430$ s (Section~\ref{sec:X-ray}), though we note that EP FXT SNe EP\,250108a and EP\,250827b both had prompt emission timescales on the order of $\sim 1000$ s \citep{Li+25, Srinivasaragavan25b}. We crudely perform this calculation assuming $t_r \sim$ 430 -- 1000 s given the uncertainty in timescale. We solve Eq. \ref{eqtime} and Eq. \ref{eqlum} in parallel, using a characteristic value of $\kappa = 0.2 \, \rm{cm^2 \, g^{-1}}$ \citep{Srinivasaragavan25b}. We do not have good constraints on the initial shock velocity, as the earliest spectrum we obtain a velocity measurement for is at $\sim 2.5$ days (see \S \ref{sec:velocities}), where $v_t \sim 0.1c$. The maximum velocity allowed for a sub-relativistic shock breakout is $\sim 0.35c$, consistent with the maximum value of the ejecta derived from the modeling of the radio upper limits along with the photospheric velocity at early times (Section~\ref{sec:redback_radio}), and just below the relativistic threshold where pair production begins to affect the emission. Thus we can comfortably assume $v_t \sim 0.15c$ -- $0.35c$. Solving Equations \ref{eqtime} and \ref{eqlum} in tandem using the above assumptions, we derive $R_{\rm{e}} \sim  2.1 \times 10^{12}$ -- $ 5.0 \times 10^{12}$ cm and $M_{\rm{e}} \sim 2 \times 10^{-7} $ -- $6.0 \times 10^{-6} \, \rm{M_\odot}$. The radius derived is consistent with the radii expected from slightly inflated Wolf-Rayet progenitors \citep{Shiode2014} and that found for XRF\,080109/SN\,2008D ($R_{\rm{e}} \sim  4 \times 10^{12}$~cm; \citealt{Soderberg+08,Svirski+14}). 

Next, we compute the expected peak energy from the SBO, to test consistency with the reported peak energy of the WXT observation of $\sim 0.16$ keV (Section~\ref{sec:X-ray}). If the radiation was fully thermalized, we expect a blackbody spectrum and a temperature that goes as
\begin{equation}
    T_{\rm{BB}} \approx \left(\frac{L_{\rm{SBO}}}{4\pi R_{\rm{e}}^2 \sigma_{\rm{SB}}}\right)^{1/4}
\end{equation}
where $\sigma_{\rm{SBO}}$ is the Stefan-Boltzmann constant. Substituting our known values into this expression, we find $T_{\rm{BB}} \sim 3.3$ -- $5.0 \times 10^5$ K, or a peak energy of 0.03 -- 0.05 keV. However, we need to ensure that the radiation is in true thermal equilibrium, and compute the thermal coupling coefficient in an expanding gas from \citet{Nakar2010} to check this, given as 

\begin{eqnarray}\label{EQ eta Def}
\nonumber% \nonumber to remove numbering (before each equation)
  \eta 
   &\approx&  \frac{7 \cdot 10^{5} {\rm ~s~}}{\min\{t,t_d\}}
    \left(\frac{\rho}{10^{-10}{\rm g/cm^{3}}}\right)^{-2} \left(\frac{kT_{BB}}{100 eV}\right) ^{7/2}  ,
\end{eqnarray}
where $k$ is the Boltzmann constant, and ${\min\{t,t_d\}}$ is the minimum time between the characteristic timescale and the diffusion time. The characteristic timescale here is the end of the prompt emission timescale of $t \sim 430$ -- $1000$ s. If $\eta < 1$, then the approximation of thermal equilibrium holds, and the observed temperature $T = T_{\rm{BB}}$. However, if $\eta > 1$, then emission will be dominated by free-free processes, and the spectrum resembles a Wien spectrum that is also affected by Comptonization of photons by neighboring electrons \citep{Nakar2010}. We compute $\eta$, substituting our known values along with taking into account the density increases by a factor of seven due to compression of the shock \citep{WaxmanSBO}. Through this calculation, we derive $\eta \sim 4$ -- $141$, for the range of densities derived earlier. Therefore, the approximation of thermal equilibrium does not hold. Indeed, free-free emission will dominate, and the spectrum can be represented by a Wien spectrum characterized by a temperature
\begin{eqnarray}
    T{\xi(T)^2} = T_{\rm{eq}}{\eta^2}
\label{eq21}
\end{eqnarray}
where $\xi(T)$ is the Comptonization correction factor, given by 
\begin{equation}\label{EQ xi}
  \xi(T) \approx \max\left\{1,\frac{1}{2}\ln[y_{max}]\left(1.6+\ln[y_{max}]\right)\right\},
\end{equation}
where $y_{\rm{max}}$ is the Compton parameter, 
\begin{equation}\label{EQ ymax}
    y_{max} \equiv \frac{kT}{h\nu_{min}}=3  \left(\frac{\rho}{10^{-9}~{\rm g/cm^{-3}}}\right)^{-1/2} \left(\frac{T}{\rm 100
    eV}\right)^{9/4}. 
\end{equation}
We solve Equation \ref{eq21} for $T$, substituting our known values, and find $T \sim 0.15$ -- $3.2$  keV, consistent with the value found from the WXT detection. We therefore conclude that analytic arguments support an SBO origin for EP\,260321a. 

\subsection{\textsc{Redback} Evidence for SBO and Constraints on CSM}
\label{sec:redback_radio}

Having established that EP\,260321a is analytically consistent with SBO, we next use \textsc{Redback} to test whether the X-ray and radio data are consistent with a compact SBO origin. The public X-ray data provides only sparse temporal information. We therefore treat the X-ray modeling as a consistency check rather than as a unique inference of the CSM structure. We further use the radio data to constrain the circumstellar environment at larger radii. 

For the X-rays, we fit the WXT and FXT luminosities with an SBO-like model~\citep{Fryer+26} that computes the fading X-ray luminosity from shocked
material emerging from an effective emitting area. The free parameters are the injected energy, $E_{\rm in}$, the slope of the high-velocity ejecta tail, the maximum Lorentz factor, $\Gamma_{\rm max}$, the emitting area, $A$, the local density, $\rho$, the radiative efficiency, $f_{\rm eff}$, and an emergence
time, $t_{\rm emerge}$. \rr{Specifically, The effective emitting area, $A$, parameterizes the total surface area of the shocked material contributing to the X-ray emission. It is related to an effective radius by $R_{\rm eff} = (A/\pi)^{1/2}$, which can represent either a full spherical breakout surface ($A = 4\pi R^2$) or a smaller, patchy or aspherical region if the breakout is geometrically confined. This could be the case for a clumpy CSM or a preferential viewing direction through a low-opacity channel. Values of $R_{\rm eff} \ll R_{\rm progenitor}$ would therefore suggest an aspherical or partial breakout. $\Gamma_{\rm max}$ is the maximum Lorentz factor of the ejecta, setting the high-energy cutoff of the velocity distribution and the characteristic velocity of the fastest-moving material. For a sub-relativistic shock, $\Gamma_{\rm max} \lesssim$ a few, while $\Gamma_{\rm max} \sim 10–100$ is appropriate for relativistic or mildly relativistic breakouts. The parameter $p$ describes the power-law index of the Lorentz factor distribution of the fastest-moving shocked ejecta, $dN/d\Gamma \propto \Gamma^{-p}$. Steeper values of $p$ concentrate the energy in slower (lower-$\Gamma$) material, while shallower values allow a significant high-velocity tail. This determines how rapidly the X-ray luminosity declines after the shock emerges, since higher-$\Gamma$ ejecta radiate more energetically but exhaust their energy faster.}

% We note that $\Gamma_{\rm max}$ is the velocity of the fastest moving ejecta, which, assuming homologous expansion will be a small fraction of the bulk ejecta, that travels at the characteristic velocity probed by the optical models.) 
\rr{We note that $\Gamma_{\rm max}$ characterizes the velocity of the fastest-moving, outermost ejecta responsible for the X-ray SBO emission. This fast component constitutes a negligibly small mass fraction compared to the bulk ejecta, whose characteristic expansion velocity, $v_{\rm ej}$, is independently constrained by the optical SN models (Section \ref{sec:redback_sn}).}
This parameterization is not identical to the analytic variables in Section~\ref{sec:analytic_constraints}; in particular, the emitting area can represent a patchy or aspherical breakout surface rather than the full spherical radius as was previously assumed.

The SBO model (Appendix~\ref{fig:xray_space}) reproduces the rapid decline from WXT to FXT and prefers $\log_{10}(E_{\rm in}/{\rm erg})=47.47$ with a 90\% credible interval of $46.68$--$50.03$, $\Gamma_{\rm max}\simeq19$ with a broad 90\% interval of $2.3$--$80$, and $t_{\rm emerge}\simeq200$~s. 
The inferred effective emitting radius, $R_{\rm eff}=(A/\pi)^{1/2}$, is $1.5\times10^{10}$~cm, with a broad 90\% interval of $1.0\times10^{9}$--$6.4\times10^{11}$~cm. This radius is somewhat smaller than that from XRF\,080109 \citep{Soderberg+08,Svirski+14}. However,  given the limited and preliminary public X-ray data we do not conclude they are inconsistent at present. The posterior on density is also broad, with a median $\rho\simeq4.5\times10^{-8}~{\rm g~cm^{-3}}$. 
Given this limited X-ray information, these should not be taken as strong constraints but rather as further support (when combined with the analytical calculation) towards establishing the SBO origin of EP\,260321a. In particular, this supports that a compact, rapidly cooling SBO component can account for the X-ray luminosity evolution without requiring a luminous, long-lived relativistic afterglow.

The radio data probe a different physical scale: synchrotron emission from the fastest SN ejecta interacting with the outer wind-like CSM. We model this component in \textsc{Redback} using the \texttt{synchrotron\_massloss}
model~\citep{Chevalier2017}. For simplicity, we assume $\epsilon_e=\epsilon_B=0.1$, a range on 
$p=2.5$--$3.0$, allow shock velocities
$v_s=3\times10^4$--$1.5\times10^5~{\rm km~s^{-1}}$ ($\simeq0.1$--$0.5c$), and fit the wind mass-loading parameter $\dot{M}/v_w$ for which $\dot{M}$ is the mass-loss rate and $v_w$ is the velocity of the wind. The posterior is again an allowed-region constraint because the radio data are upper limits (Figure~\ref{fig:radio_constraints}). We find
$v_s=5.3\times10^4~{\rm km~s^{-1}}$ with a 90\% interval of $3.1\times10^4$--$1.1\times10^5~{\rm km~s^{-1}}$ ($\beta\simeq0.18$, with a 90\% interval of $0.10$--$0.36$), and
\begin{equation}
    \log_{10}\left[\frac{\dot{M}/v_w}
    {M_\odot~{\rm yr^{-1}}/({\rm km~s^{-1}})}\right]
    \leq -7.92 .
\end{equation}
\rr{We show the allowed light curves from our space compared to the full span of light curves allowed by our prior ranges in Appendix Figure~\ref{fig:corner_massloss}, highlighting that the posterior distributions are data-driven.}. In addition to the radius derived from the X-ray SBO signal, that EP\,260321a's SN counterpart is of Type Ic-BL further favors a Wolf-Rayet progenitor (e.g., \citealt{WoosleyHeger06}). For a Wolf-Rayet-like wind velocity of $v_w=1000~{\rm km~s^{-1}}$ \citep{Cappa+04}, the median mass-loss rate corresponds to $\dot{M}\lesssim1.2\times10^{-5}~M_\odot~{\rm yr^{-1}}$. This is consistent with the $\dot{M} \approx 10^{-5}~M_\odot~{\rm yr^{-1}}$ inferred from radio detections of SN\,2008D \citep{Soderberg+08,vanderhorst+11}, and observed mass-loss rates in Wolf-Rayet stars \citep{Cappa+04}.
The fastest allowed ejecta require lower mass loading: conditioning on
$v_s>0.15c$ gives a 95th-percentile upper limit of $\log_{10}[(\dot{M}/v_w)/(M_\odot~{\rm yr^{-1}}/{\rm km~s^{-1}})]\lesssim-7.77$.

Taken together, the X-ray and radio constraints favor a picture in which the X-rays arise from a compact SBO component close to the progenitor, while the radio limits require that any faster ejecta encounter only modest mass-loading at larger radii. This is consistent with SBO from a compact Wolf-Rayet progenitor or a small amount of nearby material, and disfavors an ordinary on-axis GRB afterglow (e.g., \citealt{PanaitescuKumar02,KangasFruchter21}). It does not exclude aspherical structure, clumpy CSM (e.g., \citealt{Fryer+20}), or an off-axis jet, which require later-time radio observations to test robustly. Overall, we find that the inferred properties of EP\,260321a's progenitor are broadly consistent with those of XRF\,080109/SN\,2008D.

\section{Analysis of SN\,2026gzf}
\label{sec:sn2026gzf}

\subsection{Comparison to SESNe Light Curves}
\label{sec:comp}

Having established that EP\,260321a was likely the product of SBO, we next investigate the properties of its Type Ic-BL counterpart, SN\,2026gzf, and contextualize these with SESNe. First, we compare SN\,2026gzf's light curve to the range of SESNe discovered with and without high-energy counterparts (FXTs, GRBs). In Figure~\ref{fig:phot_comp} we plot SN\,2026gzf's rest-frame $r$-band light curve against optically discovered SNe Ic-BL \citep{Srinivasaragavan+24}, optically discovered Type Ib/c SNe \citep{Drout+11}, GRB SNe \citep{Clocchiatti+11,Matheson+03,Cano+11,Olivares+12,Levan+14_ULGRBs,Greiner+15,Perley+14,Cano+17,Izzo+19,Hu+21,Rastinejad+24} and previous FXT SNe (including SBO SN\,2008D; \citealt{Soderberg+06,Mirabal+06,Modjaz+06,Soderberg+08,Modjaz+09,SCG+25,Rastinejad+25,Srinivasaragavan+25}). For the GRB and FXT SNe, we set $\delta t = 0$ to be the time of the GRB or FXT trigger and do not remove emission from the synchrotron afterglow or early shock cooling. For the optically discovered Type Ic-BL and Ib/c SNe we normalize the light curve peak times to the rest-frame peak time of SN\,2026gzf. We then linearly interpolate the $r$-band light curves and determine the 1$\sigma$ confidence interval on the range of Type Ic-BL and Ib/c light curves. We correct all light curves for Galactic extinction in the direction of the burst \citep{SchlaflyFinkbeiner11}. We further correct SN\,2008D and SN\,2010bh for notable host extinction along the line-of-sight \citep{Soderberg+08,Starling+11}.

\begin{figure}
\centering
\includegraphics[width=0.49\textwidth]{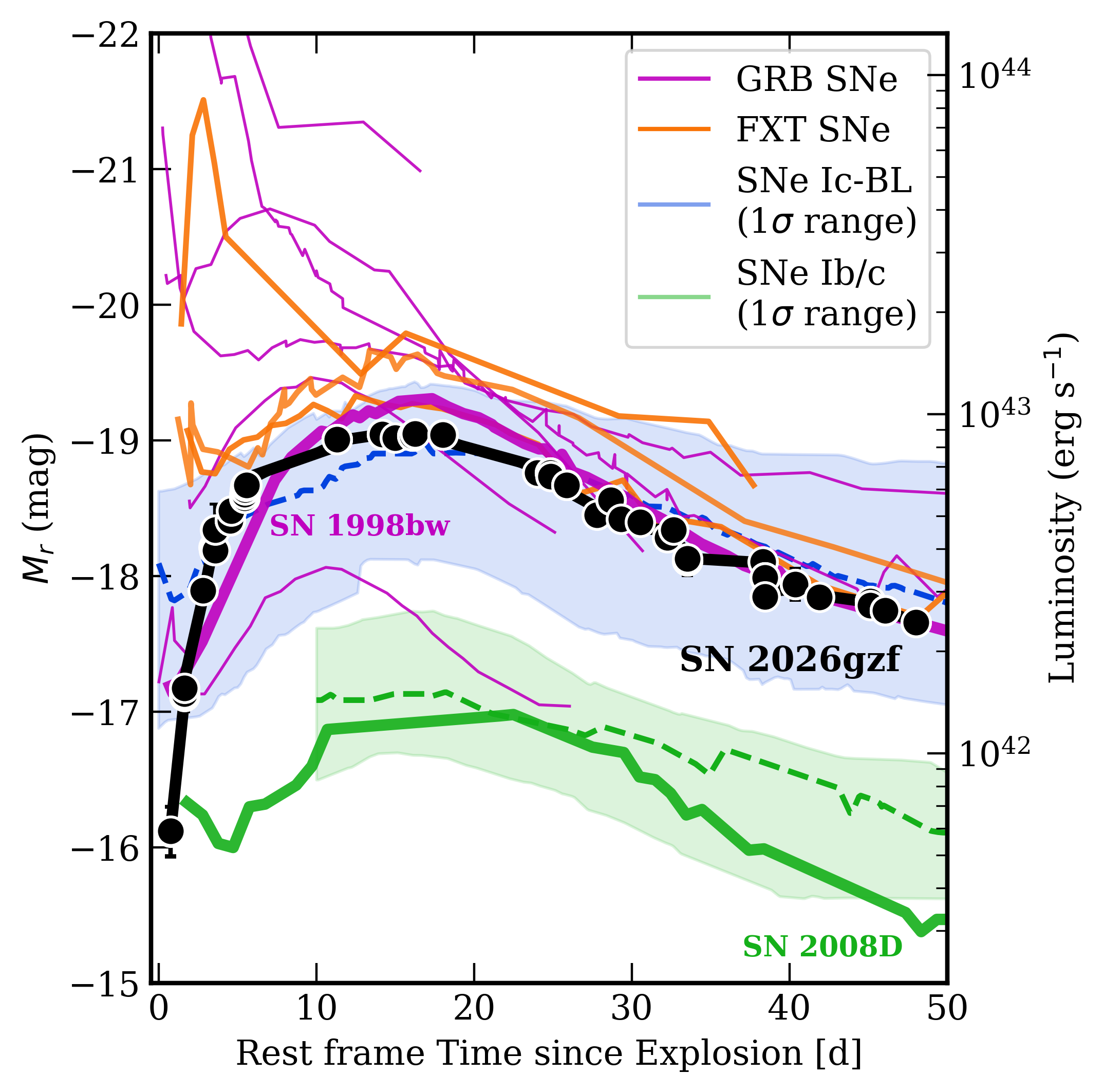}
\caption{Comparison of rest-frame $r$-band light curves for SN\,2026gzf (black) against GRB (magenta; including synchrotron afterglow emission) and FXT (orange), optically-discovered Type Ic-BL (blue; \citealt{Srinivasaragavan+24}), Ib/c (green; \citealt{Drout+11}) SNe. We plot the median (dashed line) and 1$\sigma$ range (shaded regions) of Types Ib/c and Ic-BL SNe. We highlight GRB SN\,1998bw and Type Ib SBO SN\,2008D with thick solid lines.}
\label{fig:phot_comp}
\end{figure}

SN\,2026gzf is well within the 1$\sigma$ range of Type Ic-BL SN $r$-band light curves and consistent with the range of light curves of GRB and FXT SNe (Figure~\ref{fig:phot_comp}). Its peak magnitude of $M_{r} \sim -19.1$~mag is more luminous than 75\% of optically discovered Type Ic-BL SNe \citep{Taddia2019,Srinivasaragavan+24}. Compared to the median Type Ic-BL light curve, SN\,2026gzf exhibits a more rapid rise at $\delta t \lesssim 1$~week. However, not all Type Ic-BL are well-sampled at these early epochs \citep{Srinivasaragavan+24}. Further, GRB and FXT SNe are often contaminated by afterglow emission on these timescales. Finally, we highlight that SN\,2026gzf is $\approx$10 times more luminous compared to the Type Ib SBO SN\,2008D, though this is in keeping with the medians of their respective SN types (Figure~\ref{fig:phot_comp}).

\subsection{Expansion Velocities}
\label{sec:velocities}

We next measure the velocity of the $\lambda$5169 \AA~ Fe II line for our spectra of SN\,2026gzf taken with GMOS, NGPS and SOAR. The relative blueshift of the middle of the broad Fe II absorption trough is a well-tested proxy for the photospheric expansion velocity ($v_{\rm{ph}}$) of the SN \citep{Modjaz+16}. We use the open source  SESNspectraLib\footnote{https://github.com/metal-sn/SESNspectraLib} \citep{Modjaz+16, Liu2016} package to smooth each spectrum and use SESNspectraPCA\footnote{https://github.com/metal-sn/SESNspectraPCA} to calculate the blueshift of the Fe II line relative to standardized Type Ic spectroscopic templates at similar phases. We calculate the total uncertainty by adding the uncertainty on the velocity of the mean Type Ic template in quadrature with the uncertainty on the relative blueshift. 

We show our expansion velocity measurements of SN\,2026gzf in Figure~\ref{fig:sne_icbl_props} against other FXT, GRB, Type Ic-BL and Ic SNe \citep{Iwamoto+98,Mazzali+00,Mazzali+02,Mazzali+03,Mazzali+06,Pian+06,Sauer+06,Mazzali2008,Modjaz+09,Taddia2019,Corsi+23,Anand+24,Srinivasaragavan+24,Srinivasaragavan25b}. Overall, SN\,2026gzf falls well within the range of expansion velocities for SESNe. At our earliest epoch ($\delta t = 2.5$~days), SN\,2026gzf expanded at a greater velocity ($v_{ph} = 37,700^{+10600}_{-5200}$ km s$^{-1}$) compared to FXT SN\,2008D and the handful of Type Ic-BL SNe observed at a similar epoch. However, we note that this temporal region is poorly sampled for most event classes, in particular for GRB SNe due to afterglow contamination.

\begin{figure*}
\centering
\includegraphics[width=\textwidth]{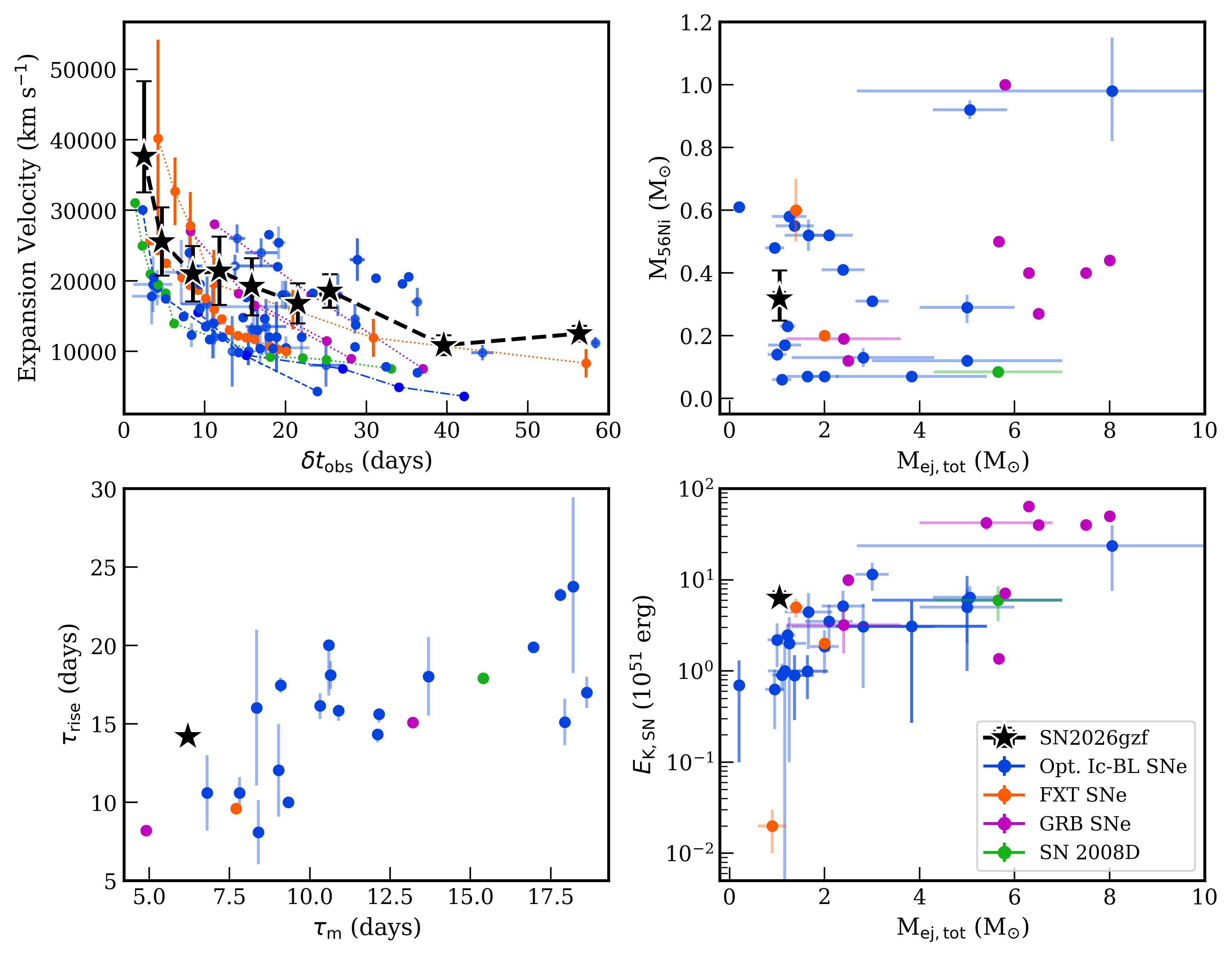}
\caption{Comparison of the properties of SN\,2026gzf derived from our analysis (black stars; Section~\ref{sec:sn2026gzf}) with GRB (magenta), FXT (orange; including Type Ib SN\,2008D which is shown in green), and optically discovered Type Ic-BL SNe (blue). In the top left panel we plot the SN expansion velocities (Section~\ref{sec:velocities}; \citealt{Iwamoto+98,Mazzali+00,Mazzali+02,Mazzali+03,Mazzali+06,Pian+06,Sauer+06,Mazzali2008,Modjaz+09,Taddia2019,Corsi+23,Anand+24,Srinivasaragavan+24,Srinivasaragavan25b}). In the right top and bottom panels we compare SN\,2026gzf's $M_{\rm ej}$, $M_{\rm Ni}$, and $E_{K}$ inferred from our CSM + $^{56}$Ni model (Section~\ref{sec:csm_ni56_model}). In the bottom left panel we plot the time from explosion to peak ($\tau_{\rm rise}$) against the diffusion timescale ($\tau_m$), highlighting that SN\,2026gzf's diffusion timescale is relatively fast compared to other events \citep{Lyman+16,Srinivasaragavan+24}. Overall, SN\,2026gzf's properties are most consistent with optically discovered Type Ic-BL SNe.}
\label{fig:sne_icbl_props}
\end{figure*}

\subsection{Near-IR Search for He Absorption}
\label{sec:he_limit}

The amount He present in the ejecta of Type Ic-BL SNe, and thus the extent to which He stripping occurs prior to core-collapse, is currently poorly constrained by observations. When correcting for blueshifting using our Fe II velocities (Section~\ref{sec:velocities}), we do not observe significant absorption at the expected positions of He I (e.g., $\lambda\lambda\lambda\lambda 4471, 5876,6678, 7061$) in our optical spectra. 

\begin{figure}
\centering
\includegraphics[width=0.49\textwidth]{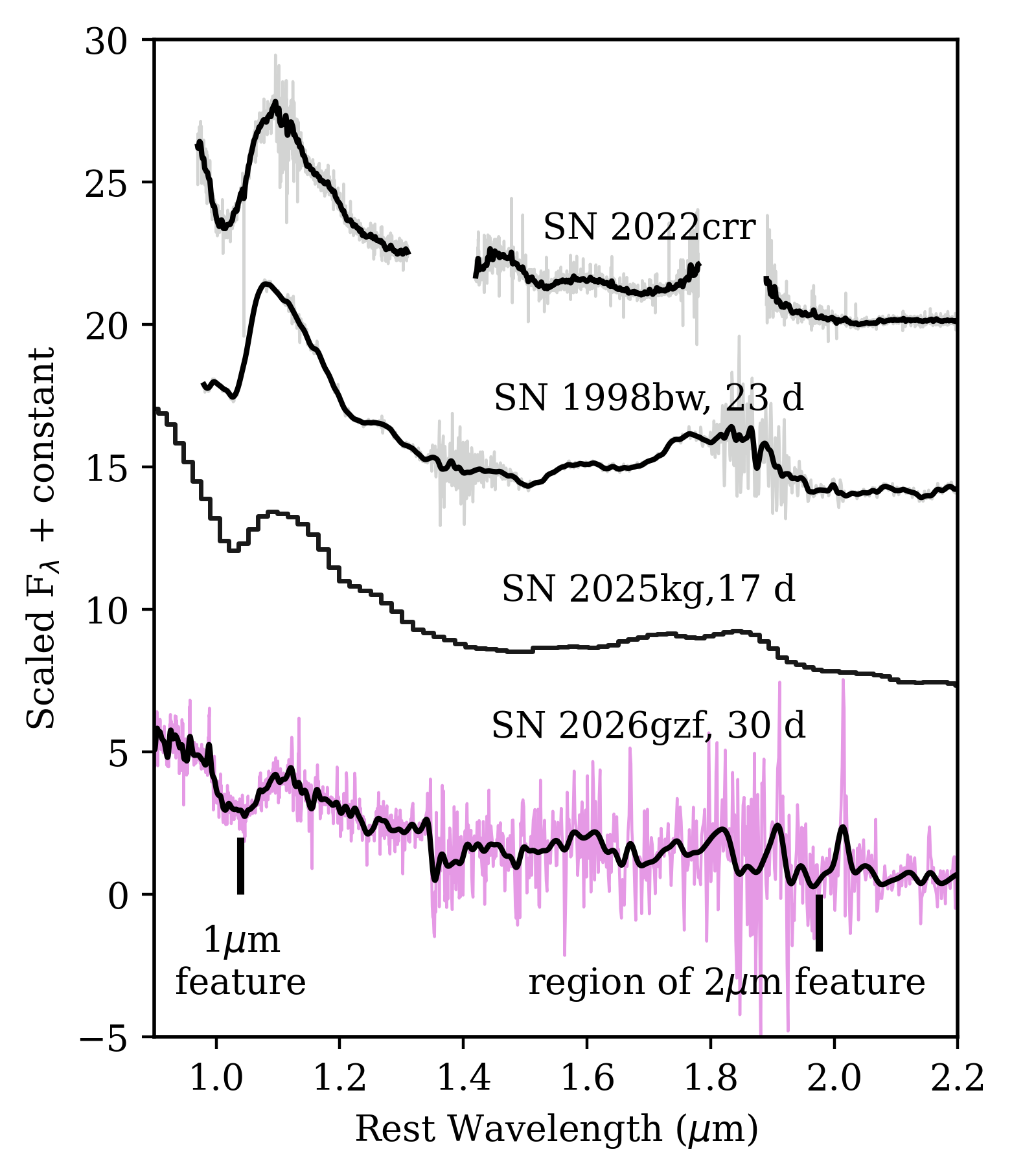}
\caption{Magellan/FIRE near-IR spectrum of SN\,2026gzf (bottom), compared to near-IR spectra of past Type Ic-BL SNe \citep{Patat+01,Tinyanont+24,Rastinejad+25}. We mask regions contaminated by telluric features. While we observe an absorption feature $\sim$1~$\micron$, the expected 2$\micron$ feature is not well-detected in our FIRE spectrum due to low signal-to-noise, precluding a firm detection of He I.}
\label{fig:nir_spec}
\end{figure}

We search for He absorption in our Magellan/FIRE near-IR spectrum of SN\,2026gzf, taken at $\delta t = 30$~days. He produces stronger absorption features in the near-IR ($\lambda$1.083~$\mu$m and $\lambda 2.058~\mu$m) compared to the optical lines. However, due to the nearby C I $\lambda 1.0693$~$\mu$m and Mg II $\lambda 1.0927$~$\mu$m lines, the detection of absorption at $\sim \lambda$1~$\mu$m is difficult to ascribe to He I without full radiative transfer calculations. He I $\lambda 2.058~\mu$m is comparatively less contaminated, but often poorly constrained due to low signal-to-noise or overlap with Telluric lines.

In Figure~\ref{fig:nir_spec} we compare the near-IR spectrum of SN\,2026gzf to that of Type Ic-BL SNe SN\,2025kg (associated with EP\,250108a; \citealt{Rastinejad+25}), SN\,1998bw (associated with GRB\,980425; \citealt{Patat+01}) and SN\,2022crr (no known high-energy counterpart; \citealt{Tinyanont+24}). We observe an absorption feature at $\sim 1 \mu$m for SN\,2026gzf, in keeping with the near-IR spectra of previous SESNe \citep{Patat+01,Tinyanont+24,Rastinejad+25,Schneider+26}. However, we do not observe significant absorption at the expected location of the He I $\lambda 2.058~\mu$m feature. Due to low signal-to-noise in the region, we cannot rule out that this absorption may be present. Robust constraints on the presence of He I absorption necessitates full radiative transfer calculations (e.g., \citealt{Shahbandeh+22,Lu+26}), outside the scope of this work.

\subsection{\textsc{Redback} SN Modeling}
\label{sec:redback_sn}

We model the Galactic extinction-corrected, well-sampled optical and near-IR light curve of
SN\,2026gzf with \textsc{Redback} \citep{Sarin+24} to infer its explosion
parameters and to test the presence of a heating channel beyond some centrally concentrated radioactive material. We explore five models to fit the light curve of SN\,2026gzf: a
one-zone Arnett-like $^{56}$Ni model, a $^{56}$Ni mixing model, a magnetar model, a pure CSM-interaction model, and a combined CSM-interaction and $^{56}$Ni model. For all fits, we use the nested sampler \texttt{Pymultinest} \citep{pymultinest} through \texttt{Bilby} \citep{ashton19}. The likelihood includes both detections and upper limits, with the latter treated using the \textsc{Redback} Gaussian upper limit likelihood. Unless otherwise specified, we use broad, uninformative uniform or log-uniform priors for all parameters. 

All models are fit in observed-frame magnitude space, and assume an initially expanding photosphere that begins to recede once the ejecta temperature drops to $T_{\rm floor}$. The spectral energy distribution (SED) is assumed to be a blackbody. This is a useful approximation for fitting the broad-band light curve, but it does not model line blanketing, spectral features, or deviations from a single thermal photosphere. These general limitations are applicable to all models we consider. 

Throughout, we fix the redshift to $z=0.0343$ and include host-galaxy extinction, $A_V$, as a nuisance parameter using a Fitzpatrick extinction law with $R_V=3.1$ \citep{Fitzpatrick1999}. 
For the CSM models, we also run physically constrained fits
using the standard \textsc{Redback} CSM constraints: the CSM photosphere must
sit between the inner and outer CSM radii, and the CSM diffusion time must be
shorter than the shock-crossing time. These constraints are critical to preventing CSM configurations that are mathematically allowed by the luminosity prescription but physically
self-inconsistent. We fix the optical opacity to $\kappa=0.07~{\rm cm^2~g^{-1}}$ for all models apart from the $^{56}$Ni mixing fit,
where the model includes an internal temperature-dependent opacity
parameterization. This value is typical of Type Ic-BL SN modeling~\citep[e.g.,][]{Srinivasaragavan2024}. 

The model light curve credible regions are shown in Figure~\ref{fig:sne_models}. Overall, we find that the combined CSM+$^{56}$Ni model gives the strongest physical and statistical support among the models explored. As we will discuss in Section~\ref{sec:discussion}, the model also provides a natural self-consistent picture for the multi-wavelength properties and the progenitor of this event. We emphasize that the following modelling does not fold in any information from our radio or X-ray analysis above as well as any spectra. Instead, it is simply exploring how well we can fit the SN lightcurve from a data and physical perspective. We compare our constraints on CSM properties to those from the EP X-ray emission and radio non-detections in Section~\ref{sec:comb_csm_picture}.

\begin{figure*}
    \centering
    \includegraphics[width=0.85\textwidth]{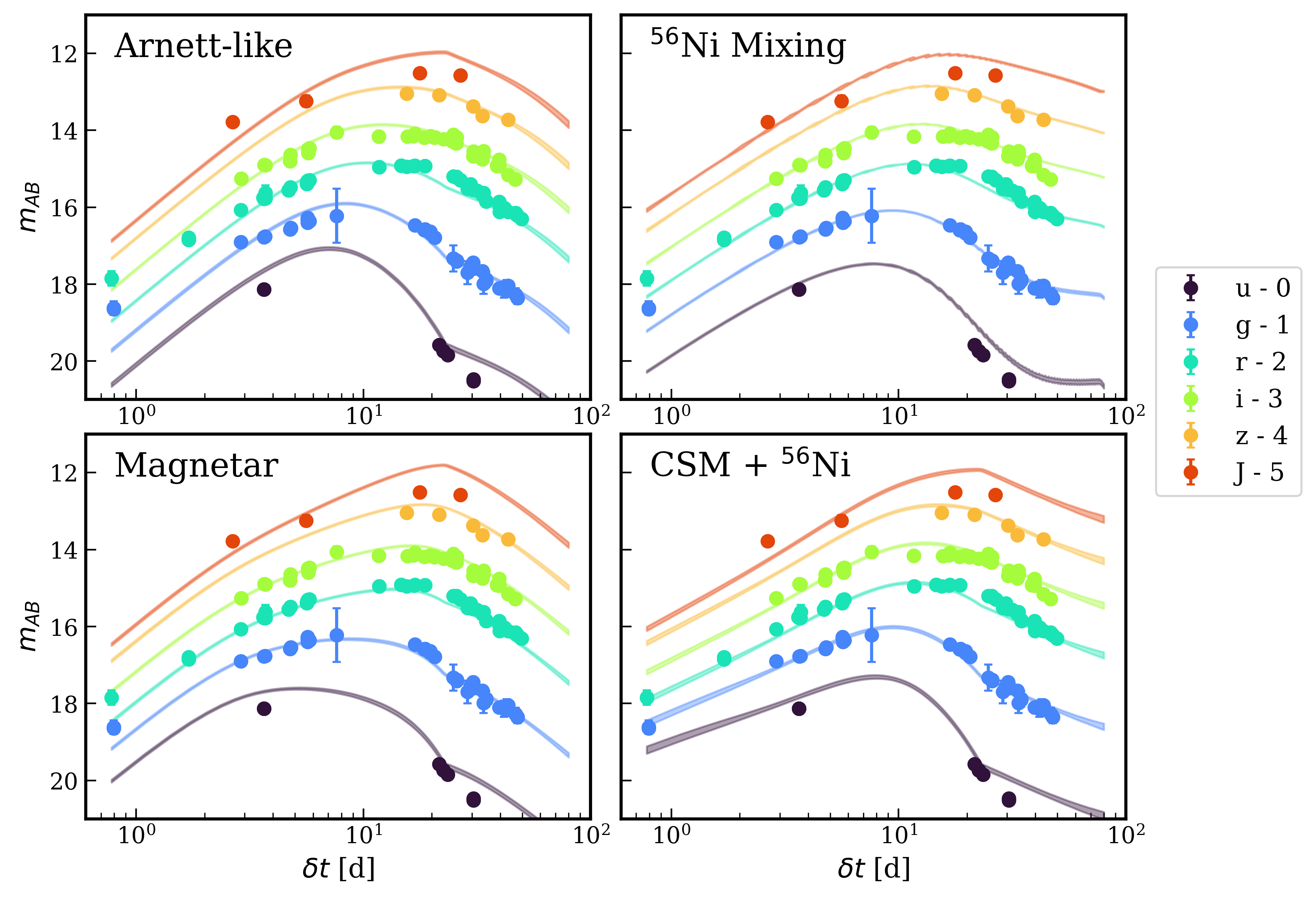}
    \caption{We fit the $ugrizJ$-band light curves of SN\,2026gzf with five
    models in \textsc{Redback} (the pure CSM-interaction model is not shown here as it requires infeasible $M_{\rm ej}$ and $E_{K}$; Section~\ref{sec:csm_interaction_model}). We disfavor the one-zone Arnett-like,
    magnetar, and pure CSM-interaction models because they either provide poor
    statistical fits or require nonphysical parameters
    (Sections~\ref{sec:arnett}, \ref{sec:magnetar_model}, and
    \ref{sec:csm_interaction_model}). The combined CSM+$^{56}$Ni model is our
    preferred model. Relative to the $^{56}$Ni mixing model, the constrained
    CSM+$^{56}$Ni fit improves the Bayesian evidence by
    $\Delta\ln Z \simeq 47$.}
    \label{fig:sne_models}
\end{figure*}

\subsubsection{One-Zone Arnett Model}
\label{sec:arnett}

We first model the light curve using a one-zone Arnett-like model \citep{Arnett82,Valenti+08}. This model assumes that the radioactive heating from $^{56}$Ni and $^{56}$Co decay is a single zone and the raw radioactive energy diffuses through ejecta with a constant grey opacity ($\kappa$). The free parameters are the ejecta mass, $M_{\rm ej}$, the $^{56}$Ni mass fraction, $f_{\rm Ni}$, the ejecta
velocity, $v_{\rm ej}$, the gamma-ray opacity, $\kappa_\gamma$, the
temperature floor, $T_{\rm floor}$, and the host extinction, $A_V$. 

In particular, the raw radioactive decay luminosity is processed through the semi-analytic diffusion solution~\citep{Arnett82,Pinto+00}. We also compute the characteristic photon diffusion timescale,
\begin{equation}
   \tau_m^2 = \frac{2\kappa_{\rm opt} M_{\rm ej}}{\beta c v_{\rm sc}},
\end{equation}
where $\beta\simeq13.8$ is the diffusion constant and $v_{\rm sc}$ is the
characteristic ejecta velocity. $\gamma$-ray leakage is controlled by $\kappa_\gamma$. The model assumes a single characteristic $v_{\rm ej}$ and homologous expansion; and a single zone of material. 

To reproduce SN\,2026gzf's fast rise and luminous peak at fixed $\kappa$, a fit to the light curve requires either a low effective diffusion mass, a high $v_{\rm ej}$, or high fraction of radioactive material. Nevertheless, the one-zone model provides a valuable benchmark for comparison to population studies of SNe (e.g., \citealt{Cano+17_review,Taddia2019,Srinivasaragavan+24}) and for providing estimates of the explosion properties from the light curve.

Applying the model to SN~2026gzf, we find that the one-zone model cannot reproduce the rapid optical rise without pushing to an unphysical radioactive composition. The posterior gives $M_{\rm ej}=0.48\pm0.01~M_\odot$ and
$f_{\rm Ni}=0.991^{+0.007}_{-0.014}$, corresponding to
$M_{\rm Ni}=0.473\pm0.012~M_\odot$. Although the resulting kinetic energy is
not extreme, $E_K\simeq2.6\times10^{51}$~erg, the inferred ejecta would be almost entirely $^{56}$Ni. This is not physically plausible for a Type Ic-BL SN. The Bayesian evidence is also the lowest of the models considered
($\ln Z=-304.7$). Thus, we disfavor the one-zone model on both physical and statistical grounds.

\subsubsection{$^{56}$Ni Mixing Model}
\label{sec:mixing_model}

The failure of the one-zone model and SN\,2026gzf fast and bright rise naturally motivate a model in which a fraction of the radioactive material is mixed into the outer ejecta. 
Such mixing is supported by numerical simulations~\citep[e.g.,][]{Khatami2019,REichert+23}, and may naturally occur in some Type Ic-BL due to jet-driven turbulence~\citep{Grimmett+21,REichert+23}. We utilize the $^{56}$Ni mixing model~\citep{Sarin2026_mixing} in \textsc{Redback}. 
In addition to the parameters employed in the Arnett-like model (Section~\ref{sec:arnett}), the $^{56}$Ni mixing model assumes that the ejecta mass distribution follows a radial power law, with $^{56}$Ni deposited in layers up to the free parameter $f_{\rm mix}$, as well as a temperature-dependent $\kappa$ parameterization designed to follow results from numerical simulations \citep{Nagy2018}. The model divides the ejecta into velocity shells and deposits the $^{56}$Ni into the inner fraction of the ejecta specified by $f_{\rm mix}$; larger values place radioactive material into faster, lower optical-depth layers. The default ejecta profile is a broken power law with a shallow inner density slope and a steep outer density slope. Each shell is evolved with radioactive heating, $\gamma$-ray leakage, adiabatic losses and a local diffusion time before the shell luminosities are summed. Though semi-analytic, this model captures the key physical effect missing from the one-zone model: radioactive heating can emerge earlier if some $^{56}$Ni is mixed outward.

The mixing model provides a substantially better description of the data than the one-zone model, with $\ln Z=-125.7$. The inferred ejecta parameters are also reasonable for a SESNe explosion:
$M_{\rm ej}=1.11^{+0.63}_{-0.23}~M_\odot$,
$E_{\rm SN}=4.39^{+2.50}_{-0.90}\times10^{51}$~erg, and
$M_{\rm Ni}=0.398^{+0.005}_{-0.004}~M_\odot$. The model prefers strong mixing, $f_{\rm mix}=0.80^{+0.11}_{-0.13}$, consistent with SN\,2026gzf's rapid rise, and overall captures the
light curve evolution. However, the model fails to fully account for the early blue/UV data, which could be indicative of a failure of our blackbody SED assumption.

\subsubsection{Magnetar Model}
\label{sec:magnetar_model}

We also fit the light curve with a generalized magnetar model~\citep{sarin+22,Omand+24} implemented in~\textsc{Redback}. This model injects magnetar spin-down power into the ejecta and diffuses the resulting luminosity through an expanding photosphere. Critically, this model captures the coupling between the magnetar spin-down energy and the dynamical evolution of the ejecta; a robust outcome of simulations~\citep{Suzuki+21} which dramatically alters the predictions of magnetar-driven SNe~\citep{Omand+24}. The free engine parameters are the initial spin-down luminosity, $L_0$, the spin-down timescale, $t_{\rm sd}$, and the braking
index, $n$, which could be fixed to $n=3$ if one assumes that the magnetar only spins down due to vacuum dipole radiation~\citep{Ostriker+69}. We sample with broad uninformative priors on all parameters but impose $M_{\rm ej}>0.1~M_\odot$ to avoid
solutions that are non-physical for a Type Ic-BL SN.

The magnetar model is statistically viable and outperforms the $^{56}$Ni
mixing model, with $\ln Z=-104.3$. However, the posterior is pinned to the lower ejecta-mass boundary, $M_{\rm ej}=0.101~M_\odot$ with a 90\% credible interval of $0.100$--$0.104~M_\odot$, and has a very short effective diffusion time, $\tau_m\simeq1.8$~days. 
The inferred characteristic velocity is $v_{\rm ej}\simeq2.7\times10^4~{\rm km~s^{-1}}$, while the magnetar engine has $L_0\simeq1.1\times10^{43}~{\rm erg~s^{-1}}$ and a broad spin-down timescale posterior centered at $t_{\rm sd}\sim1.6\times10^3$~d. We note that the braking index is essentially unconstrained (largely by the lack of late-time data that provides a constraint on this quantity), but assuming $n=3$, our parameters would imply a magnetar with an initial spin-period of $p_{0}\sim 4$~ms and a magnetic-field of $B_{p} \sim 2\times10^{13}$~G, which are not similar to the typical properties for putative magnetar's in Type Ic-BL SNe~\citep{Zhu2026}. Therefore, although the magnetar model can provide a good fit the data, it requires an extremely low effective ejecta mass and unusual birth properties. Thus, we disfavor the magnetar model on physical grounds.

\subsubsection{CSM Interaction Model}
\label{sec:csm_interaction_model}

We next fit a model powered purely by interaction between the SN ejecta and a shell of CSM. The implementation follows the semi-analytic CSM-interaction formalism of previous works \citep{Chatzopoulos+13, Villar+17_model,Jiang+20}, as implemented in \textsc{Redback}. 

The model assumes homologously expanding ejecta with a broken power-law density profile interacting with a spherical CSM density profile
$\rho_{\rm CSM}=q r^{-s }$, where $q=\rho {r_0}^{s }$. The forward and reverse shock luminosities are computed from the analytic self-similar solution with a constant efficiency in converting kinetic energy to thermal radiation, which we fix to $0.5$. The free parameters are $M_{\rm ej}$, $v_{\rm ej}$, $T_{\rm floor}$, $A_V$, the CSM mass, $M_{\rm CSM}$, the density normalization, $\rho$, the inner radius of the CSM shell, $r_0$, and the CSM density-profile index, $s$. The total CSM mass, $M_{\rm CSM}$, together with $r_0$, $\rho$, and $s $, sets the outer radius of the shell. The CSM is assumed to be static, i.e., $v_{\rm csm} \ll v_{\rm ej}$, such that the dynamical evolution of the CSM can be ignored, outside of the interval $r_0 \leq R \leq r_{\rm out}$, the CSM density is set to zero. 
The diffusion calculation depends most directly on the optically-thick CSM mass, where \rr{we define} the mass exterior to the CSM photosphere \rr{as characterized by} $\kappa>2/3$. Thus $M_{\rm CSM}$ and the outer radius is typically weakly constrained unless the SN ejecta have fully traversed the shell.

Sampling with broad uninformative priors, we find that an unconstrained pure CSM model can reproduce much of the light curve shape
($\ln Z=-118.2$), but only by requiring an implausibly massive and energetic ejecta component: $M_{\rm ej}=26.4^{+2.5}_{-4.3}~M_\odot$ and
$E_K\simeq1.1\times10^{53}$~erg. This is inconsistent with expectations for a Type Ic-BL SN. Imposing stricter constraints on the ejecta mass and additionally physical constraints such that the CSM photosphere lies within the shell and the diffusion time is shorter than the shock-crossing time, the pure CSM fit \rr{cannot accommodate for both the late-time brightness and the late-time decay, resulting in a poor fit to the data} ($\ln Z=-292.8$). The combination of infeasible parameters and physical inconsistency disfavor a pure interaction-powered model.

\subsubsection{CSM Interaction and $^{56}$Ni Model}
\label{sec:csm_ni56_model}

Finally, we fit a model that combines CSM interaction with the single-zone radioactive
$^{56}$Ni heating. This model uses the same CSM-interaction engine described above, while adding a $^{56}$Ni-$^{56}$Co decay component diffusing through the ejecta and optically thick CSM. We sample with broad uninformative priors but impose the physical constraints of shock crossing time and photosphere location as described in Section~\ref{sec:csm_interaction_model}. 

This model provides the strongest physically interpretable fit among the models considered, with $\ln Z=-78.7$. It is favored over the magnetar model by $\Delta\ln Z\simeq26$, over the $^{56}$Ni mixing model by $\Delta\ln Z\simeq47$, and over the one-zone model by
$\Delta\ln Z\simeq226$. Relaxing the physical constraint gives a higher fit, but places the CSM at much larger radii and violates the self-consistency conditions of the interaction model; we therefore use the result with a constrained prior for our physical interpretation.

Our inferred values are $M_{\rm ej}=1.05^{+0.12}_{-0.12}~M_\odot$, $f_{\rm Ni}=0.303^{+0.046}_{-0.036}$, corresponding to $M_{\rm Ni}=0.318^{+0.015}_{-0.013}~M_\odot$. The kinetic energy is $E_K=(6.30^{+0.87}_{-0.85})\times10^{51}$~erg, corresponding to a characteristic ejecta velocity of $v_{\rm ej}\simeq2.45\times10^4~{\rm km~s^{-1}}$. These parameters are all consistent with expectations of Type Ic-BL SNe. 

The CSM component is compact compared with typical late-time interaction systems but more extended than the radius one may expect from a stripped, compact star. The fit gives
$r_0=4.39^{+5.07}_{-2.63}~{\rm AU}$ or ($6.56^{+7.58}_{-3.93}) \times 10^{13}$~cm for the inner edge of the CSM shell with a density of $\rho=(4.45^{+20.2}_{-3.39})\times10^{-13}~{\rm g~cm^{-3}}$, and a nearly wind-like density profile, $s =1.98^{+0.02}_{-0.03}$. The total CSM mass is not tightly constrained due our short observing window, but we infer $M_{\rm CSM}=0.029^{+0.605}_{-0.027}~M_\odot$ at 68\% credibility. The optically thick mass participating directly in the diffusion problem is much smaller, $M_{\rm CSM,thick}\simeq1.1\times10^{-3}~M_\odot$, and the CSM photosphere lies at $R_{\rm ph,CSM}\simeq11.7~{\rm AU}$. As mentioned, the inferred outer CSM radius is broad, with a median of $\sim150$~AU, reflecting the weak leverage of the optical light curve on low-density material at large radii.

\subsubsection{Comparison of Inferred Explosion Properties}
\label{sec:best_model}

In Figure~\ref{fig:sne_icbl_props} we compare the inferred explosion properties
for the CSM + $^{56}$Ni model to those of other Type Ic-BL, FXT, and GRB SNe
\citep{Cano+17_review,Srinivasaragavan+24}. These comparison values are
mostly derived using one-zone Arnett models, therefore our comparison is not strictly model-independent, though the presence of CSM at large radii is rarely seen. Nevertheless, SN\,2026gzf lies within the broad range of $M_{\rm ej}$, $M_{\rm Ni}$, and $E_K$ inferred for Type Ic-BL SNe.

For the fiducial CSM+$^{56}$Ni model, we find $\tau_m\simeq6.2$~d. Thus, although the total time from explosion to maximum light is not exceptionally short, the inferred diffusion time is among
the shorter values for SESNe (Figure~\ref{fig:sne_icbl_props}). This is consistent with
the main conclusion of the model comparison: the early light curve cannot be
explained by a centrally concentrated one-zone $^{56}$Ni model alone, and
requires either extensive outward mixing of the radioactive material (Section~\ref{sec:mixing_model}) or an
additional early heating component (Section~\ref{sec:csm_ni56_model}). Among the models tested here, the combination of $^{56}$Ni heating and CSM interaction provides the most compelling explanation.

\section{Environmental Analysis}
\label{sec:host}

\subsection{Global Host Galaxy Properties}
\label{sec:host_props}
We next present a detailed analysis of the host galaxy of SN\,2026gzf, SDSS J095942.88+002506.2 using public photometry and spectroscopy as well as the Python-based stellar population modeling inference code \texttt{Prospector} \citep{Leja2019, jlc+2021}. We describe the details of our analysis in Appendix Section~\ref{sec:host_appendix}.

For easy interpretation of the \texttt{Prospector} results, we calculate physical properties from the sampled values. For example, we determine a stellar mass ($M_*$), mass-weighted stellar population age ($t_m$), and present-day (0--100~Myr) SFR and specific SFR (sSFR = SFR/$M_*$) using the SFH and $M_F$. We \rr{also convert the amount of light attenuated from young ($\tau_{V,1}$) and old ($\tau_{V,2}$) stars} to a total $V$-band magnitude, by multiplying their sum by 1.086.  In Table~\ref{tab:host_prop}, we list the medians and 68\% confidence intervals of these properties. 

EP\,260321a's host is a young ($t_m = 3.12$~Gyr) dwarf galaxy ($M_* = 10^{8.55} M_\odot$), with a moderate amount of star-formation (SFR = 0.06 $M_\odot$~yr$^{-1}$, log(sSFR) = -9.8~yr$^{-1}$). We further determine that the host has subsolar stellar ($\log(Z_*/Z_\odot) = -1.84$) and gas-phase ($\log(Z_\textrm{gas}/Z_\odot) = -0.63$)
metallicities. In addition to the \texttt{Prospector}-determined metallicities, we compute a gas-phase oxygen abundance (12+log(O/H)). Flux measurements and uncertainties for the observed [O II], [O III], and $H\beta$ lines are provided by DESI, along with uncertainties. We use the \citet{tremonti2004} R23-calibration method to determine 12+log(O/H), where R23 = log$_{10}$(([O II]+[O III])/H$\beta$). To incorporate the uncertainties into this measurement, we sample 1000 fluxes for each line from a Gaussian distribution and calculate 1000 12+log(O/H) values for the host. We place the median and 68\% confidence interval in Table~\ref{tab:host_prop} and determine that the host has a slightly sub-solar 12+log(O/H) metallicity of $=8.61$.

\begin{table}[ht]
    \centering
    \caption{Host Galaxy Stellar Population Properties}
    \label{tab:host_prop}
    \begin{tabular}{lc}
        \hline
        \hline
        \multicolumn{2}{c}{GLOBAL} \\
        \hline
        $\log(M_*/M_\odot)$ & $8.55^{+0.06}_{-0.06}$ \\
        SFR ($M_\odot \text{ yr}^{-1}$)    & $0.06^{+0.04}_{-0.02}$ \\
        $\log(\text{sSFR})$ ($\text{yr}^{-1}$) & $-9.8^{+0.22}_{-0.21}$ \\
        $t_m$ (Gyr)                         & $3.12^{+2.72}_{-1.32}$ \\
        $\log(Z_*/Z_\odot)$               & $-1.84^{+0.18}_{-0.12}$ \\
        $\log(Z_\textrm{gas}/Z_\odot)$               & $-0.63^{+0.09}_{-0.06}$ \\
        $A_V$ (mag)                       & $0.09^{+0.07}_{-0.05}$ \\
        $12+\log{\textrm{(O/H)}}$                     & $8.61^{+0.1}_{-0.12}$ \\
        \hline
        \multicolumn{2}{c}{KNOT} \\
        \hline
        $\log(M_*/M_\odot)$ & $6.97^{+0.94}_{-0.07}$ \\
        SFR ($M_\odot \text{ yr}^{-1}$)    & $0.16^{+0.05}_{-0.06}$ \\
        $\log(\text{sSFR})$ ($\text{yr}^{-1}$) & $-7.85^{+0.19}_{-0.9}$ \\
        $t_m$ (Gyr)                         & $1.69^{+3.37}_{-1.56}$ \\
        $A_V$ (mag)                       & $0.43^{+0.21}_{-0.15}$ \\
        \hline
    \end{tabular}
\tablecomments{Our derived stellar population properties (median and 68\% confidence intervals) for the host galaxy of SN\,2026gzf and the star-forming knot within the host.}
\end{table}

To further contextualize these results, we compare them to the host galaxies of long GRBs, optically discovered Types Ic-BL, Ib/c and II SNe. We gather host stellar masses and sSFRs for spectroscopically classified Type Ic-BL, Ibc, and II SNe at $z<0.2$ in \citet{frankenblast}, which were modeled with a nearly identical \texttt{Prospector} template as discussed here. We collect host 12+log(O/H) metallicities for these populations in \cite{qin2024}. We obtain long GRB host stellar masses and sSFRs from \citet{svensson2010, perley2013, wang2014, niino2017}, and 12+log(O/H) metallicities from \citet{wang2014}, and limit this sample to $z<0.5$ to be in a more comparable redshift range to the other host populations. We show cumulative distributions (CDFs) of the host stellar masses, sSFRs, and 12+log(O/H) metallicities in Figure~\ref{fig:sm_sfr}. We find that the host of SN\,2026gzf falls in the bottom 32\% of stellar masses for long GRB hosts, and the bottom $10\%$ for Type Ic-BL, Ibc, and II hosts. On the other hand, the host of SN\,2026gzf has a more typical sSFR when compared to the long GRB and Type Ic-BL host populations, falling in the bottom 32\% and 60\% of sSFRs, respectively. The host sSFR is much higher than observed for typical CCSN hosts, falling in the top 23\% of sSFRs for Type II SN hosts and 17\% for Type Ibc SN hosts. Finally, the host metallicity is much lower than observed from Type II (bottom 18\%) and Ibc (bottom 8\%) hosts, but more typical for long GRB (bottom 81\%) and Type Ic-BL (bottom 30\%) hosts. In summary, we find that the EP\,260321a's host is comparable to both long GRB and optically discovered Type Ic-BL SNe, though its properties are moderately more consistent with those of long GRBs.

\begin{figure*}
    \centering
    \includegraphics[width=0.95\textwidth]{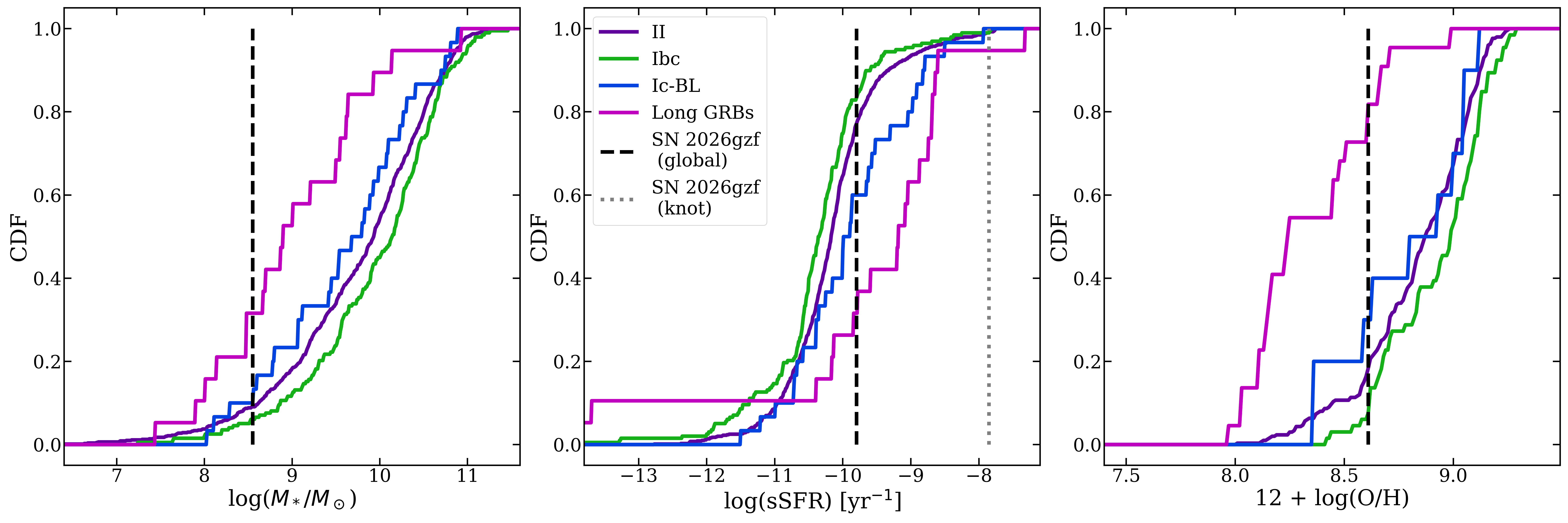}
    \caption{The global host galaxy stellar mass ($M_*$; \textit{left}), sSFR (\textit{middle}), and 12+log(O/H) metallicity (\textit{right}) of SN\,2026gzf (black dashed line) compared to the respective CDFs of the hosts of long GRBs (magenta; \citealt{svensson2010, perley2013, wang2014, niino2017}), Type Ic-BL (blue), Ibc (green), and II (dark purple) SNe \citep{qin2024,frankenblast}. We also show the sSFR of the star-forming knot coincident with the location of SN\,2026gzf (grey dotted line). The host itself is low mass, low metallicity, and star-forming, with properties that are overall more typical for long GRB hosts than the other transient host populations. The knot is extremely star-forming in comparison to the global host properties and other transient host populations.}
    \label{fig:sm_sfr}
\end{figure*}

\subsection{Stellar Population Modeling of SN\,2026gzf's Underlying Blue Knot}
\label{sec:knot_props}

We also perform a simplistic environmental analysis of the blue knot consistent with the location and redshift of SN\,2026gzf. We obtain aperture photometry of the knot from DECaLS DR10 and list the photometric measurements in Table~\ref{tab:host}. To determine its stellar populations properties, we model the photometry with the same \texttt{Prospector} model discussed in Section~\ref{sec:host_props}, without the parameters used to fit a spectrum. As we only fit four bands of optical photometry, the star formation and stellar mass are the only parameters we consider to be robustly constrained. We present our stellar population modeling results of the knot in Table~\ref{sec:host_props}. Due to the width ($\sim 1$\arcsec) and orientation of the slit used in our SN spectra, it is possible that light from both the blue knot and broader galaxy is contributing to the observed emission lines. Thus, we do not use the galaxy emission lines observed in our SN spectra in our analysis.

The properties of this knot appear to be consistent with an H II region. We determine that the knot has $M_* = 10^{6.97} M_\odot$, SFR = 0.16~$M_\odot$~yr$^{-1}$, and log(sSFR) = -7.85 yr$^{-1}$. The knot's SFH appears to be clearly rising with a rapid increase in SFR in the past 30~Myr, suggestive of a recent burst of star formation and potentially indicative that the knot is an H II region. The SFR and SFH history are consistent with those of the central region of the massive H II region, 30 Doradus in the Large Magellenic Cloud (e.g., \citealt{Cignoni+15}).

Overall, we conclude that SN\,2026gzf was born in or nearby an exceptionally star-forming environment. In comparison to the global host properties of SN\,2026gzf, the knot's sSFR is greater by $\sim 2$ orders of magnitude. Moreover, as shown in Figure~\ref{fig:sm_sfr} the knot has much higher sSFR than the global properties of the long GRB, Type Ic-BL, Ibc, and II host populations: falling in the 99\% of sSFR for the SN hosts, and the 94\% for long GRB hosts. However, we note that such local environment studies are not possible for the majority of SESNe host galaxies and thus direct comparison of the knot with the global host properties is discouraged.

\subsection{Estimate of SFR from Radio Observations}
\label{sec:radio_sfr}

Finally, we estimate a lower limit on the host-galaxy SFR required to account for the observed radio emission, using the measured integrated flux density and the following relation \cite{2011ApJ...737...67M}:
\begin{equation}
\label{eq:2}
    \left( \frac{{\rm SFR}_{\rm 6.2\,GHz}}{M_{\odot}{\rm yr}^{-1}} \right) =
    6.35 \times 10^{-29} \left( \frac{L_{\rm 6.2\,GHz}}{\rm{erg\,s}^{-1}{\rm Hz}^{-1}} \right).
\end{equation}  
From this, we calculate an SFR of $\approx $0.037\,M$_{\odot}$\,yr$^{-1}$ from our radio observations. This is a factor of $\sim 4$ below the SFR inferred for the knot from optical imaging (Table~\ref{tab:host_prop}). However, as the VLA detection is marginal ($\lesssim 5\sigma$) we estimate a large uncertainty on the radio SFR. Future, deeper VLA imaging and further high-resolution optical observation of the star-forming knot are necessary to constrain the amount of dust obscuration in the knot.

\section{Discussion}
\label{sec:discussion}

\subsection{Combined Progenitor Picture}
\label{sec:comb_csm_picture}

In Sections~\ref{sec:analysis} - \ref{sec:host}, we leverage X-ray, optical, and radio observations of
EP\,260321a/SN\,2026gzf to obtain independent constraints on the progenitor and CSM in its immediate environment. Here, we summarize our constraints and discuss the broader picture we may infer from our multi-wavelength constraints.

Our preferred optical model combines CSM interaction with $^{56}$Ni
radioactive heating. The model favors a CSM inner radius
$r_0=4.4^{+5.1}_{-2.6}$~AU ($6.6^{+7.6}_{-3.9} \times 10^{13}$~cm) and a nearly wind-like density profile, $s \simeq2$. The total CSM mass is poorly constrained, $M_{\rm CSM}=0.029^{+0.605}_{-0.027}~M_\odot$, due to the absence of late-time data in the optical light curve but may be improved with further monitoring. The inferred optically thick CSM mass is much smaller, $M_{\rm CSM,thick}\sim10^{-3}~M_\odot$, with a CSM photospheric radius of $\sim12$~AU and a broad outer-radius posterior with a median of $\sim150$~AU.

The X-ray constraints probe a smaller radial scale compared to the optical. The analytic shock breakout calculation gives $R_{\rm e}\sim2.1$--$5.0\times10^{12}$~cm ($0.14$--$0.33$~AU) and $M_{\rm e}\sim$(0.2--6)$\times10^{-6}~M_\odot$.
This radius is comparable to the outer scale of an inflated Wolf-Rayet envelope (e.g., \citealt{Shiode2014}) and is well inside the characteristic optical CSM radius  (Figure~\ref{fig:csm_diagram}). The \textsc{Redback} X-ray fit also supports a compact, rapidly fading breakout-like component, with an effective emitting radius that is small compared with the optical CSM scale, although the sparse X-ray data do not determine the geometry. We therefore interpret the X-rays and optical emission as likely probing different radial zones: the X-rays originate from SBO through a compact stellar envelope or very nearby low-mass material, while the optical light curve is shaped by interaction with more extended material at radii of several AU and beyond (Figure~\ref{fig:csm_diagram}). 

There are several caveats to this interpretation. For example, if the CSM probed by the optical emission was a smooth, spherical, optically thick structure overlying the compact X-ray breakout region, the soft X-rays would likely be absorbed and reprocessed. The apparent difference between the X-ray and optical radii may then provide evidence for a non-spherical or porous CSM geometry, clumping, or a line of sight through which the compact breakout emission can escape. The difference in radii could also be pointing towards limitations in modeling such as more structured CSM that departs from a power-law as enforced by the optical model (Section~\ref{sec:csm_ni56_model}) or asymmetries in the CSM distribution, which would change our inferred parameters and by extension constraints on the CSM. Both of these would likely require further detailed modeling with more flexible and self-consistent models jointly on the full X-ray and optical dataset~\citep[e.g.,][]{Sarin2026_csm}.

The radio limits add an important constraint on the density at larger radii. For fast \rr{($\beta = 0.10$--$0.36$; 90\% confidence interval)} SN ejecta interacting with a wind-like CSM \rr{and assuming $\epsilon_e = \epsilon_B = 0.1$ (Section~\ref{sec:redback_radio})}, the radio non-detections imply a low mass-loading parameter,
\begin{equation}
    \log_{10}\left[\frac{\dot{M}/v_w}
    {M_\odot~{\rm yr^{-1}}/({\rm km~s^{-1}})}\right]\leq -7.92 ,
\end{equation}
with even lower mass loading required for the fastest ejecta. For a Wolf-Rayet-like wind velocity of $v_w=1000~{\rm km~s^{-1}}$, this corresponds to $\dot{M}\sim1.2\times10^{-5}~M_\odot~{\rm yr^{-1}}$. Expressed as a smooth wind-equivalent mass, this mass loading corresponds to only
$\sim1.7\times10^{-6}~M_\odot$ within 30~AU and
$\sim8\times10^{-6}~M_\odot$ within 150~AU. 
These values are well below the median total CSM mass inferred from the optical model but are still consistent, especially considering the different modeling assumptions. such as the spherical, static, power-law CSM enforced by the optical model. The radio data argues against a dense, smooth, spherically symmetric wind extending to large radii. The CSM is more likely to be radially confined, clumpy, asymmetric, or characterized by a steep density drop outside the region probed by the SN light curve.  

The host galaxy environment is consistent with this picture.% and potentially offers insight to the source of the shock producing EP\,260321a. 
EP\,260321a occurred in a low-mass, subsolar-metallicity, star-forming dwarf galaxy, with global properties more similar to the hosts of long GRBs than those of optically discovered SESNe  (Section~\ref{sec:sn2026gzf}). \rr{This is a potential indicator that the core-collapse of EP\,260321a's progenitor produced some jet emission, as} stars in lower metallicity environments retain greater angular momentum as they evolve \citep{Heger+03}. \rr{Either a weak ($E_{\rm k, iso} \lesssim 10^{49}$~erg) or off-axis jet could be the source of the shock which created EP\,260321a. However, we caution that the global host properties inferred in our environmental analysis are several orders of magnitude greater than probed by the shock.} EP\,260321a's X-ray and radio properties do not appear compatible with an on-axis jet origin (e.g., Sections~\ref{sec:analysis}). Longer-term radio monitoring is key to searching for emission from an off-axis jet (e.g., \citealt{Corsi+16,Leung+23,Schroeder+25,Odwyer+26}). 

\begin{figure*}
    \centering
    \includegraphics[width=0.95\textwidth]{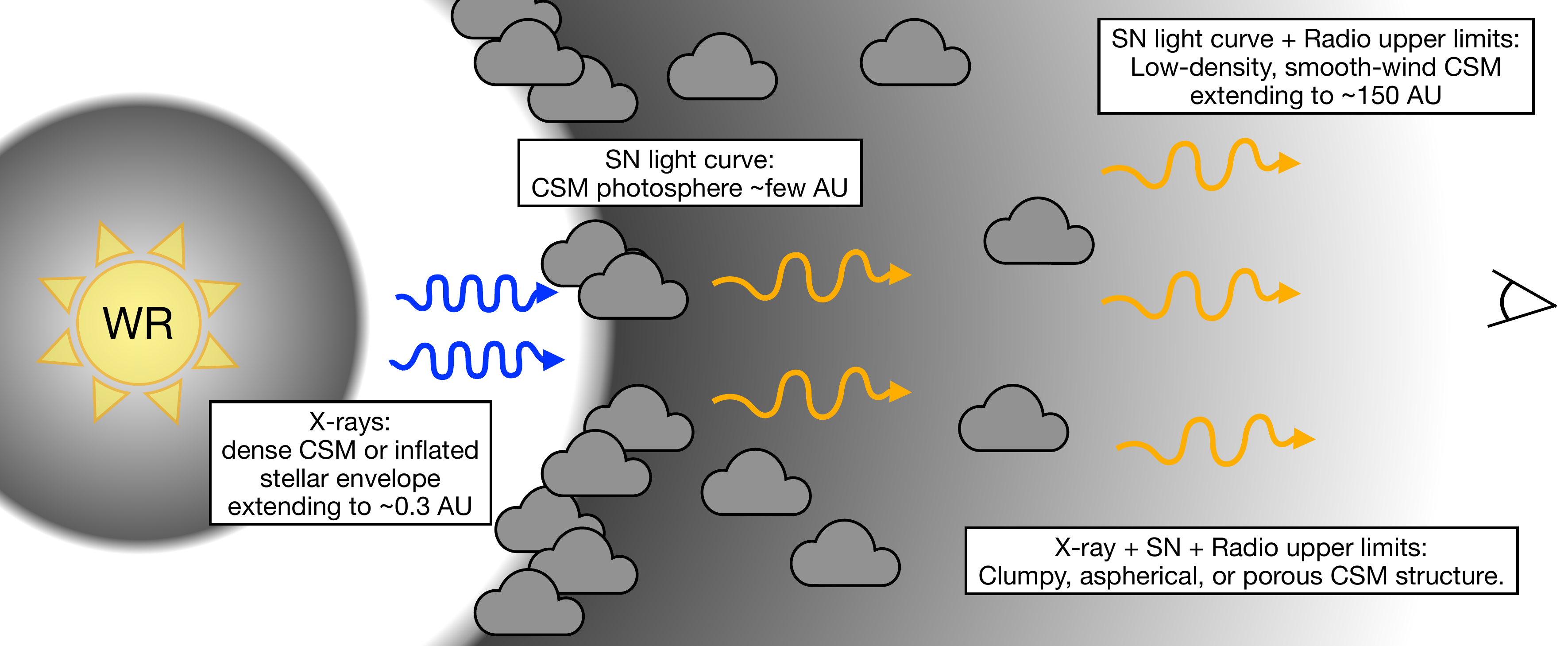}
    \caption{Cartoon diagram of the CSM orientations favored by our multi-wavelength constraints on EP\,260321a/SN\,2026gzf. Our analysis favors two distinct regions of CSM that produce the X-ray (inner region) and SN light curve (outer region). Further, we favor an outer CSM structure that is clumpy, aspherical, porous, and/or characterized by a steep density drop outside the region probed through the SN light curve.}
    \label{fig:csm_diagram}
\end{figure*}

Overall, our favored progenitor picture is a compact Wolf-Rayet-like star that underwent enhanced mass loss before explosion, producing a small amount of nearby material responsible for the X-ray breakout and additional extended, structured CSM that shaped the early optical light curve. The mass loss need not resemble a steady, smooth wind. Instead, the combined optical, X-ray, and radio constraints could be pointing towards a radially structured or asymmetric CSM configuration: compact low-mass material at $\lesssim0.3$~AU (producing the X-rays), CSM beginning at a few AU (affecting the optical light curve), and relatively low smooth-wind density at larger radii (probed through the radio; Figure~\ref{fig:csm_diagram}). A greater sample of events with early X-ray detections, dense optical follow-up, and late-time radio monitoring will be required to determine whether this CSM was produced by eruptive pre-SN mass loss, binary interaction, or the final stages of Wolf-Rayet wind evolution.

\subsection{Rates of EP\,260321a-like SBO Events}
\label{sec:rates}

While EP\,260321a-like X-ray signals are rarely detected, its optical counterpart, SN\,2026gzf, is not unusual when compared to the Type Ic-BL SNe regularly detected by optical surveys (Figure~\ref{fig:sne_icbl_props}; \citealt{Taddia2019,Srinivasaragavan+24}). We are thus motivated to consider: are the rates of optically discovered Type Ic-BL SNe consistent with EP's detection rate of X-ray SBOs? We limit our calculation to Type Ic-BL SNe, though we note that XRF\,080109's counterpart, SN\,2008D, was a Type Ib SN, for which the observed rate in optical surveys is higher (e.g., \citealt{Perley+20}). 

To determine an expected rate of X-ray SBO produced by Type Ic-BL SNe, we employ an optically discovered Type Ic-BL SNe rate confidence interval of 800 - 3000~Gpc$^{-3}$ yr$^{-1}$ (Perley et al., in prep.) from the ZTF Bright Transient Survey (BTS; \citealt{Perley+20}). This interval is drawn from integrating over the rates of Type Ic-BL SNe in peak magnitude bins brighter than $M_{r} > -16$~mag, encapsulating all Type Ic-BL's in the ZTF BTS (e.g., Figure~\ref{fig:phot_comp}; \citealt{Srinivasaragavan+24}). We next assume WXT reaches a limiting flux of $3 \times 10^{-11}$ erg s$^{-1}$ cm$^{-2}$ in its all-sky survey exposures \citep{Cheng+25}, and observes 9.3\% of the sky instantaneously with a 100\% duty cycle. Finally, we make the assumption that all FXTs from Type Ic-BL SNe SBO will reach the same peak luminosity as EP\,260321a, $L_{\rm 0.5-4~keV} = 2.2 \times 10^{44}$ erg s$^{-1}$ \citep{GCNanalysis}. Under these assumptions, we derive the horizon at which WXT will observe EP\,260321a-like FXTs of $d_L = 247$~Mpc or $z = 0.054$. Incorporating EP's sky fraction and the range of BTS Type Ic-BL SN rates results in an observed rate of $4.4 - 16$ year$^{-1}$. 

We next evaluate whether this range is compatible with EP's detection rate. EP was launched in January 2024 and completed commissioning in July 2024. Considering up the present day (May 2026) EP's survey duration is $\sim$1.8 years, within which one EP SBO has been discovered. Invoking Poisson statistics gives a 90\% (95\%) confidence interval of the observed EP SBO rate of $3 \times 10^{-3} - 4.0$ ($10^{-3} - 4.8$) year$^{-1}$. EP observations are thus consistent with the lower range of rates we infer from the BTS Type Ic-BL SNe sample at the 95\% confidence level but not at the 90\% level. Our analysis indicates some evidence that not all Type Ic-BL SNe produce EP\,260321a-like FXTs. However, we cannot confidently eliminate the possibility at present.

There are several potential factors that would lead to a lower rate inferred from EP than BTS Type Ic-BL SNe. First, there are likely systematic effects that would lead to a lower observed EP rate, including detections being interrupted by slewing, a duty cycle $<$100\%, or low signal-to-noise ratios preventing robust X-ray spectral constraints for events at the edge of the redshift threshold. Second, it is possible and indeed likely that not all Type Ic-BLs create luminous SBOs given the expected diversity in mass loss histories and explosion properties (e.g., \citealt{Fryer+26}). Whether the SN brightness and the FXT brightness are correlated remains observationally undetermined, but is supported by the two observed SBOs: the SBO XRF\,080109 was a factor of $\sim 3$ less luminous in X-rays compared to EP\,260321a and SN\,2008D's peak brightness was $\sim 10$~times less luminous compared to SN\,2026gzf. SN\,2026gzf's peak magnitude is brighter than 75\% of optically discovered Type Ic-BL SNe (Section~\ref{sec:comp}), which themselves may not be representative due to Malmquist bias \citep{Taddia2019,Srinivasaragavan+24}. If a correlation between FXT and SN luminosity exists, this would indicate most SBOs from Type Ic-BL SNe are less luminous than EP\,260321a. A larger sample of SBO events is required to constrain the diversity of observed SBO signals, their correlations with SN properties, and the rates of such events. 

\section{Conclusion}

We have presented our comprehensive optical and radio observations of the counterpart to EP's first SBO FXT: EP\,260321a. Our main conclusions are as follows:

\begin{enumerate}
    \item Analytic arguments and \texttt{Redback} modeling of EP\,260321a favor an SBO origin for the X-ray emission, representing the first X-ray SBO detected by EP.
    \item Our radio upper limits rule out an on-axis jet of $E_{\rm iso} \gtrsim 10^{49}$~erg \rr{for $n > 10^{-2}~{\rm cm}^{-3}$ and $\epsilon_e = \epsilon_B = 0.1$.}. Further, the radio places deep constraints on CSM at larger radii, favoring modest mass-loading at larger radii.
    \item EP\,260321a was accompanied by a Type Ic-BL SN, SN\,2026gzf, at $z = 0.0343$, rendering it the most nearby EP FXT known to date and the first Type Ic-BL SN with an X-ray SBO counterpart. 
    \item SN\,2026gzf's $r$-band light curve, expansion velocities, $M_{\rm ej}$, $M_{\rm Ni}$, and $E_{K}$ are consistent with the ranges for both GRB SNe and optically discovered Type Ic-BL SNe.
    \item SN\,2026gzf rapidly rose in brightness until $\delta t \sim 1$~week ($\tau_m = 6.2$~days). We show that this rapid rise cannot be explained solely with a one-zone SN model in which $^{56}$Ni radioactive decay is concentrated in the center of the explosion. Of the five models we explore, a model which incorporates CSM-interaction and $^{56}$Ni radioactive decay is favored on both physical and statistical grounds.
    \item SN\,2026gzf's host galaxy mass, sSFR and (12+log(O/H)) metallicity are within the 95\% confidence intervals of both GRB SNe hosts and optically discovered Type Ic-BL SNe hosts. The stellar mass and metallicity of SN\,2026gzf's host are more compatible with the medians for GRB SNe.
    \item Taken together, our observations and analysis favor a Wolf-Rayet progenitor star surrounded by either an inner region of dense CSM or an extended stellar envelope at $\sim 0.3$~AU. Our SN analysis also favors an additional CSM component beginning at $\sim 4$ AU which is likely clumpy, aspherical or otherwise porous. 
    \item The rates of EP SBO events are consistent at the 95\% confidence level, but not the 90\% confidence level, with the range of rates implied if all Type Ic-BL SNe produce similar signals to EP\,260321a. Continued observations from EP will determine whether these rates remain consistent.
\end{enumerate}

Our observations and analysis of EP\,260321a and SN\,2026gzf reveal significant insight to the circumstellar environment of a stripped-envelope, massive star just prior to collapse. The advent of the UV survey monitor, ULTRASAT \citep{ultrasat}, which will be sensitive to spectrally softer SBO signals, along with continued observations from EP, promise significant opportunities for multi-wavelength follow-up of SBO events in the coming decade.

\section*{Acknowledgments}

The authors thank Mery Ravasio and Eric Burns for helpful discussion regarding the manuscript.

JCR was supported by NASA through the NASA Hubble Fellowship grant \#HST-HF2-51587.001-A awarded by the Space Telescope Science Institute, which is operated by the Association of Universities for Research in Astronomy, Inc., for NASA, under contract NAS5-26555. NS is supported by the Kavli Foundation. JKL acknowledges support from the University of Toronto and Hebrew University of Jerusalem through the University of Toronto - Hebrew University of Jerusalem Research and Training Alliance program. The Dunlap Institute is funded through an endowment established by the David Dunlap family and the University of Toronto. AEN acknowledges support through the Villar Astro Time Lab from the David and Lucile Packard Foundation, the Research Corporation for Scientific Advancement (through a Cottrell Fellowship), the National Science Foundation under AST-2433718, AST-2407922 and AST-2406110, as well as an Aramont Fellowship for Emerging Science Research. SA is supported by an LSST-DA Catalyst Fellowship; this publication was thus made possible through the support of Grant 62192 from the John Templeton Foundation to LSST-DA. SA also gratefully acknowledges support from Stanford University, the United States Department of Energy, and a generous grant from Fred Kavli and the Kavli Foundation. MWC acknowledges support from the National Science Foundation with grant numbers PHY-2117997, PHY-2308862 and PHY-2409481. IA is supported by the National Science Foundation award AST 2505775, NASA grant 24-ADAP24-0159, Scialog award SA-LSST-2024-102a, and the Discovery Alliance Catalyst Fellowship Mentors award 2025-62192-CM-19. G.S. and A.Y.Q.H. acknowledge support from a Sloan Research Fellowship (Award Number FG-2024-21320) from the Alfred P. Sloan Foundation, and a Scialog Award from the Research Corporation for Science Advancement. 

FDC acknowledges support from the DGAPA/PAPIIT grant IN113424. DF's contribution to this material is based upon work supported by the National Science Foundation under Award No. AST-2401779. WJ-G\ is supported by NASA through Hubble Fellowship grant HSTHF2-51558.001-A awarded by the Space Telescope Science Institute, which is operated for NASA by the Association of Universities for Research in Astronomy, Inc., under contract NAS5-26555. AS acknowledges support from the Knut and Alice Wallenberg Foundation through the ``Gravity Meets Light" project.

Based on observations obtained at the international Gemini Observatory, a program of NOIRLab, which is managed by the Association of Universities for Research in Astronomy (AURA) under a cooperative agreement with the National Science Foundation on behalf of the Gemini Observatory partnership: the National Science Foundation (United States), National Research Council (Canada), Agencia Nacional de Investigaci\'{o}n y Desarrollo (Chile), Ministerio de Ciencia, Tecnolog\'{i}a e Innovaci\'{o}n (Argentina), Minist\'{e}rio da Ci\^{e}ncia, Tecnologia, Inova\c{c}\~{o}es e Comunica\c{c}\~{o}es (Brazil), and Korea Astronomy and Space Science Institute (Republic of Korea). Data was processed using the Gemini DRAGONS (Data Reduction for Astronomy from Gemini Observatory North and South) package.

The National Radio Astronomy Observatory and Green Bank Observatory are facilities of the U.S. National Science Foundation operated under cooperative agreement by Associated Universities, Inc.

Based on observations obtained with the Samuel Oschin Telescope 48-inch and the 60-inch Telescope at the Palomar Observatory as part of the Zwicky Transient Facility project. ZTF is supported by the National Science Foundation under Award \#2407588 and a partnership including Caltech, USA; Caltech/IPAC, USA; University of Maryland, USA; University of California, Berkeley, USA; Cornell University, USA; Drexel University, USA; University of North Carolina at Chapel Hill, USA; Institute of Science and Technology, Austria; National Central University, Taiwan, and the German Center for Astrophysics (DZA), Germany. Operations are conducted by Caltech's Optical Observatory (COO), Caltech/IPAC, and the University of Washington at Seattle, USA.

SED Machine is based upon work supported by the National Science Foundation under Grant No. 1106171. The Gordon and Betty Moore Foundation, through both the Data-Driven Investigator Program and a dedicated grant, provided critical funding for SkyPortal.

The Liverpool Telescope is operated on the island of La Palma by Liverpool John Moores University in the Spanish Observatorio del Roque de los Muchachos of the Instituto de Astrofisica de Canarias with financial support from the UK Science and Technology Facilities Council.

This paper includes data gathered with the 6.5 meter Magellan Telescopes located at Las Campanas Observatory, Chile. 

This material is based upon work supported in part by the National Science Foundation through Cooperative Agreements AST-1258333 and AST-2241526 and Cooperative Support Agreements AST-1202910 and 2211468 managed by the Association of Universities for Research in Astronomy (AURA), and the Department of Energy under Contract No. DE-AC02-76SF00515 with the SLAC National Accelerator Laboratory managed by Stanford University. Additional Rubin Observatory funding comes from private donations, grants to universities, and in-kind support from LSST-DA Institutional Members. This research uses services or data provided by the Rubin Science Platform at NSF-DOE Vera C. Rubin Observatory, which is jointly funded by the U.S. National Science Foundation and the U.S. Department of Energy, Office of Science.

The authors acknowledge the use of Google Gemini to generate plotting scripts. The authors take full responsibility for the content.

\bibliographystyle{aasjournalv7}
\bibliography{refs}

\newpage
\section*{Appendix}

\subsection{Photometric Data Reduction}
\label{sec:phot_red}

\subsubsection{LSST Photometry Prior to EP\,260321a}

% Kendall's close look at LSST photometry.
% \subsection{Potential Detection of Precursors by LSST}
%% Feel free to move this wherever you think it fits/ make it shorter! I tend to be a bit of a verbose writer. - Kendall
LSST first detected a transient at the location of SN 2026gzf on January 18, 2026 (LSST-AP-DO-314003014107006318), approximately 3 months prior to EP's detection of the SN. Repeated variability was detected from 19-24 February. However, the faintness of the detections \citep[all near LSST's single-exposure detection limit of $\sim 25$th magnitude;][]{Ive19} and the clear dipole features in the difference imaging cast doubt on the interpretation that these were true SN precursors. We perform an independent analysis of the science images released with these LSST detections \citep[LSST-AP-DS-314003014107006318 and LSST-AP-DS-170032901818679622 through 170265994707599374, downloaded from ALeRCE;][]{For21} using Scarlet2, a non-parametric forward-modeling code which makes use of machine learning priors and can model multi-resolution images from different instruments, representing both static and variable sources \citep{Ward25, Mel26}.\\ 

With Scarlet2, we extract a light curve of the transient source directly from the science images, independent of the host galaxy light. We also fit a WCS shift between the science images when necessary, eliminating any coordinate offset issues. We force the flux of the transient source to be $0$ in its dimmest epoch in each band (based on the LSST light curve provided on ALeRCE), allowing us to constrain whether the pre-SN detections are consistent with a single flux value (the value in the dimmest, `reference' epoch) or not. We find that all of the LSST transient alerts prior to the SN are consistent with no variation $\geq 5 \sigma$ in g or r bands, and the vast majority are consistent with no variation $\geq 3\sigma$. The most significant change in the r band is a $4.7 \sigma$ detection of variability between the observation on $\rm{MJD} \, 61058.3$ and the reference epoch on $\rm{MJD} \, 61095.2$. The most significant change in the g band is a $4.4 \sigma$ detection of variability between the observation on $\rm{MJD}\, 61094.2$ and the reference epoch on $\rm{MJD}\, 61095.1$. While we cannot confidently say precursors are detected for SN 2026gzf, this result shows the potential of LSST to identify SN precursor activity.\\

\subsubsection{SEDm Imaging}

We obtained $ugri$-band imaging with the Rainbow Camera mounted on the Palomar 60-inch telescope (PI: Ahumada). We reduce the data using the standard FPIPE imaging pipeline.

\subsubsection{Liverpool Telescope IO:O}

Liverpool Telescope (LT) observations were taken with the IO:O optical camera in the Sloan Digital Sky Survey (SDSS) $griz$ filters. The exposures were reduced by the automatic LT pipeline; same filter exposures per epoch in each filter were then processed with custom image subtraction and photometry software$\footnote{\url{https://github.com/kryanhinds/subphot_pipe}}$ following methods described in (\citealt{Fremling+16}), utilizing the software \texttt{SWarp} \citep{swarp}, \texttt{SExtractor} \citep{Bertin1996}, and \texttt{PSFex} \citep{Bertin_2013_psfex_ascl.soft01001B}, using reference images and calibration stars from PanSTARRS-1 \citep{Chambers+16, chambers2019panstarrs1surveys}.

\subsubsection{WINTER}

We observed ZTF26aaonmha/SN\,2026gzf in the J- and shortened H-bands with the Wide-field Infrared Transient explorer \citep[WINTER;][]{2020SPIE11447E..9KL, 2025arXiv251216753F}, a camera mounted on the 1-m telescope at Palomar Observatory. Data were reduced using the standard WINTER pipeline built using \texttt{mirar} \citep{mirar}. Astrometric calibration was performed using Gaia positions \citep{Gaia2021}, while photometric calibration was performed using 2MASS \citep{Twomass}.

\subsection{Spectroscopic Data Reduction}
\label{sec:spec_red}

\subsubsection{GMOS Spectra}

We obtained GMOS spectra using the B480 grating at a central wavelength of 520~nm and a slit width of 1''. We perform bias correction, flat-fielding, and wavelength calibration using DRAGONS \citep{DRAGONS19}. We flux calibrate our spectra using the standard star LTT6248 with DRAGONS.

\subsubsection{NGPS Spectra}

The Palomar 200-inch NGPS observation of ZTF26aaonmha/SN\,2026gzf was carried out using a 1.5” slit width and a 900 seconds exposure time. The airmass was 1.29 at the time of obseration. On-chip spatial binning was set to 2 and spectral binning was set to 3. Resolving power for this setup is R$\sim1500$. Data were reduced following standard procedures for long-slit spectroscopic data reduction using a custom pipeline developed for NGPS. Wavelength calibration was performed against daytime ThAr and FeAr arc exposures. Flux calibration was performed with an observation of the spectrophotometric standard Feige34 taken within a few hours of the science observation at airmass $\approx1.0$.

\subsubsection{Goodman Spectra}

% We obtained SOAR spectra \textcolor{red}{describe set up. Describe data reduction.}
We obtained two epochs of longslit spectroscopy of  ZTF26aaonmha/SN\,2026gzf with the Goodman High throughput Spectrograph (GHTS; \citealt{Clemens2004}) mounted on the Southern Astrophysical Research (SOAR) telescope on 22 April 2026 and 30 April 2026. The observations consisted of 3 $\times$ 600 seconds of exposures. Both observations were taken with a grating of 400 lines/mm and a 1.0'' wide slit mask in the M1 spectroscopic setup (hereafter 400M1) with $2 \times 2$ binning using the GHTS Red Camera. The 400M1 spectra cover a wavelength range of 3800 -- 7040 $\AA$.

The spectra were reduced using \texttt{pypeit} \citep{pypeit:joss_arXiv, pypeit:zenodo}, using arcs taken immediately before and/or after target observation and calibration images from the same night. Flux calibration was performed using standard stars observed on the night of the observations with an identical 400M1 setup and $2 \times 2$ binning. 

\subsubsection{FIRE Spectra}

We obtained one epoch of spectroscopy with the Folded port InfraRed Echellete (FIRE) spectrograph mounted on the Magellan Baade telescope. Observations were performed in the high throughput prism mode ($\mathrm{R}_J\sim 500$) using a 1.0\arcsec slit using high detector gain (1.3 e-/DN). We conducted 2.5 ABBA dither sequences with 137\,s per exposure, using the Sample-Up-The-Ramp (SUTR) readout mode. 

The spectrum was reduced using FIREHose v2.0 \citep{firehose}, a modified version of the IDL-based FIREHose pipeline, which relies on \texttt{spextool}. 

\subsection{Constraints from X-ray and Radio Observations}
\label{sec:radio_appendix}

\rr{We show the posterior light curves and/or corner plots from our \textsc{Redback} constraints on the properties of an on-axis jet with $\epsilon_B = 0.1$ (Figure~\ref{fig:corner_jet}) and allowed to vary over $10^{-4} \leq \epsilon_B \leq 0.1$ (Figure~\ref{fig:corner_jet_epsBfree}) as well as on the wind mass-loss rate using the X-ray data and radio upper limits (Table~\ref{tab:vla_observations}).}

\begin{figure*}
    \centering
    \includegraphics[width=0.6\textwidth]{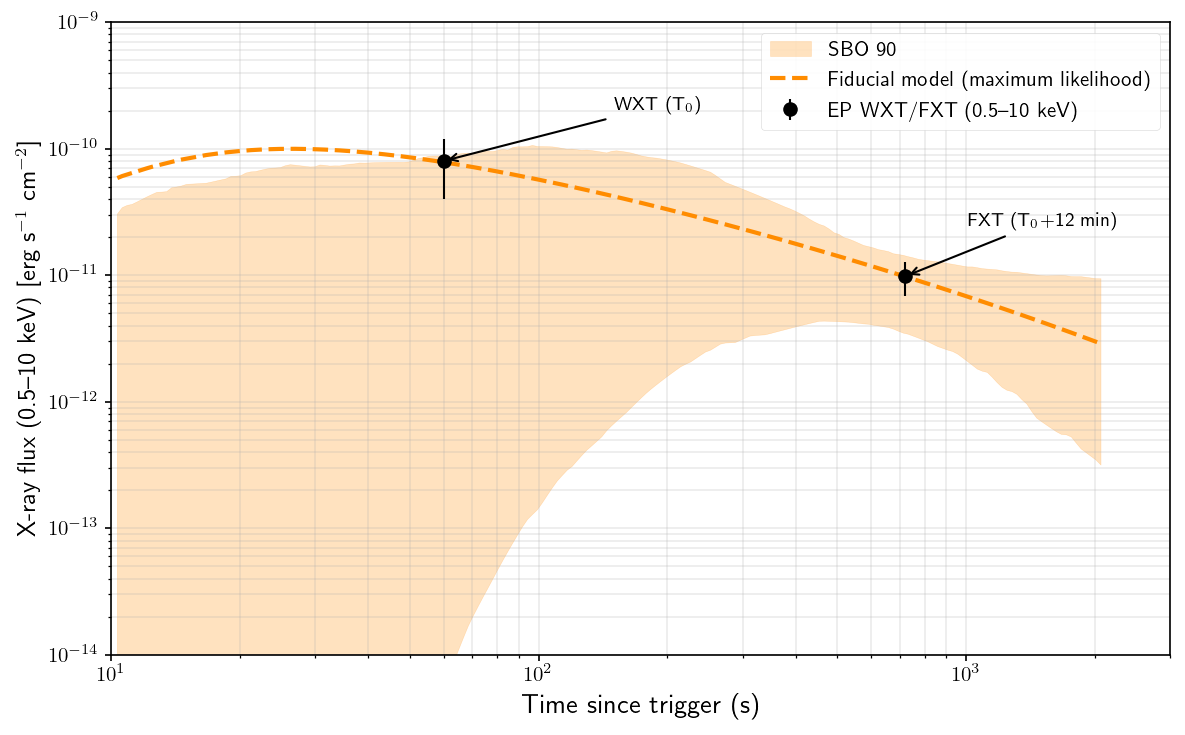}
    \caption{The allowed 90\% X-ray light curves for shock breakout emission from our fit to the EP data (black).}
    \label{fig:xray_space}
\end{figure*}

\begin{figure*}
    \centering
    \includegraphics[width=0.95\textwidth]{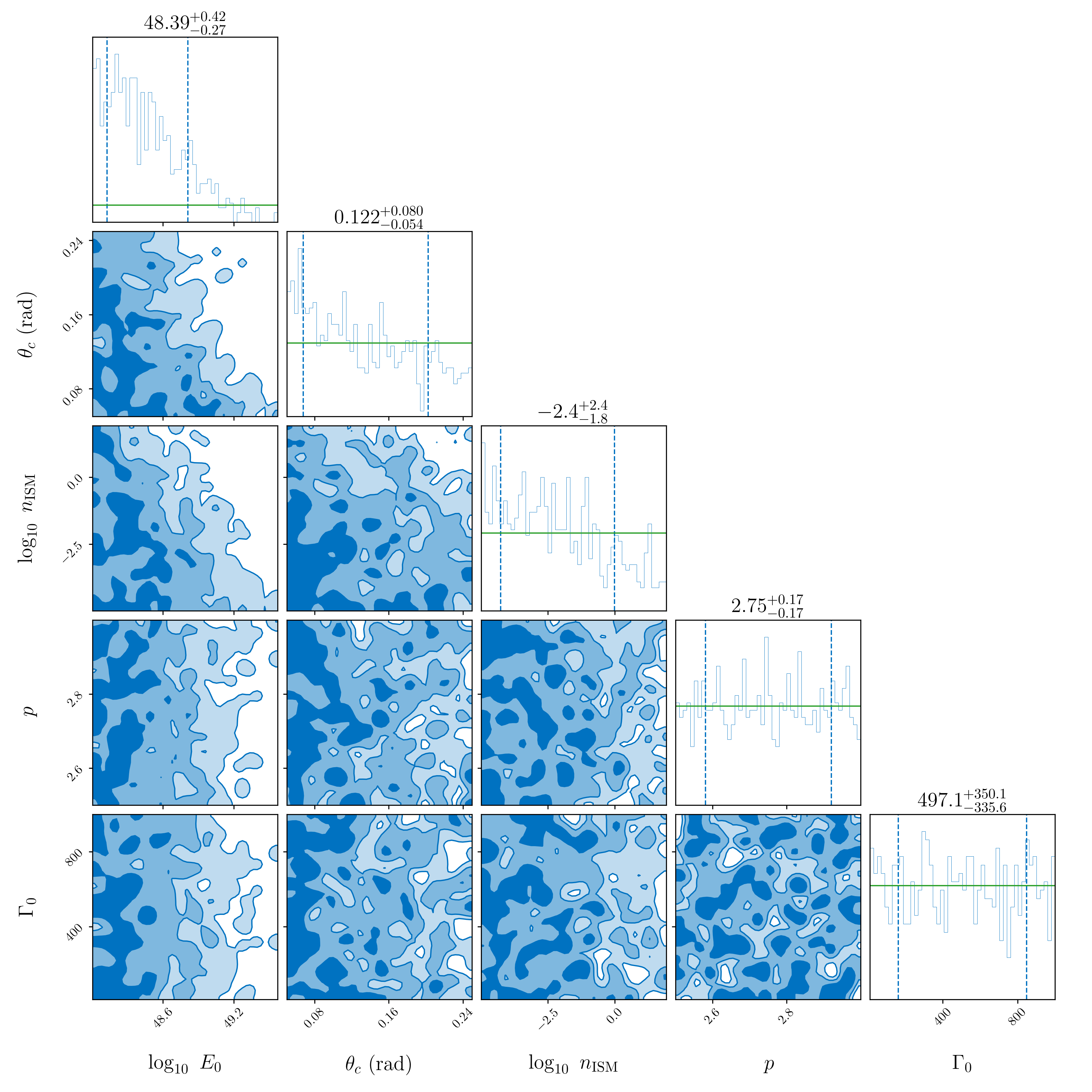}
    \caption{Corner plot showing our constraints on an on-axis jet assuming $\epsilon_e = \epsilon_B = 0.1$. The prior distribution is shown in green, highlighting that our constraint on $E_{\rm K,iso}$ is data-driven rather than prior-driven.}
    \label{fig:corner_jet}
\end{figure*}

\begin{figure*}
    \centering
    \includegraphics[width=0.95\textwidth]{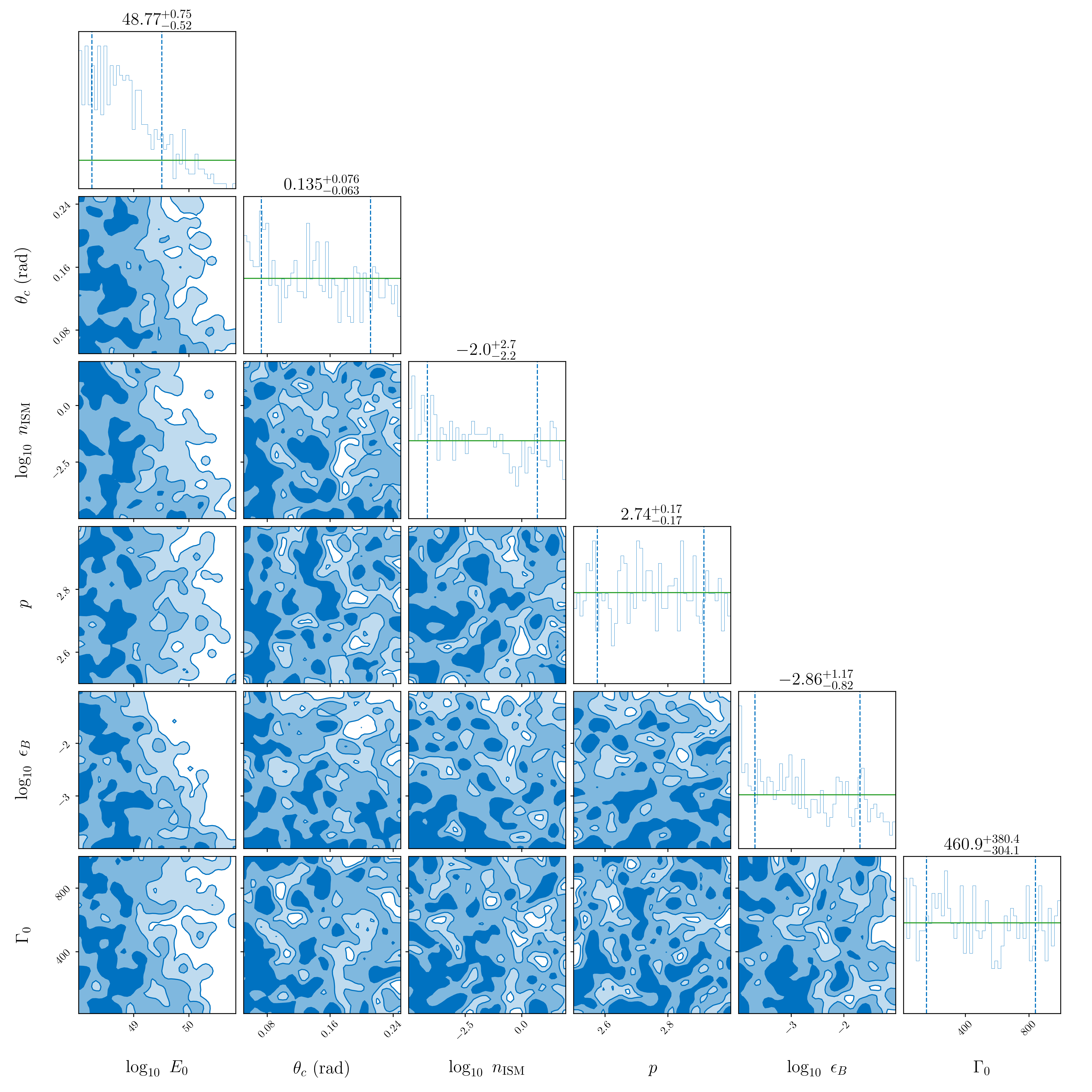}
    \caption{Corner plot showing our constraints on an on-axis jet allowing $\epsilon_B$ to vary over the range $10^{-4} \leq \epsilon_B \leq 0.1$. The prior distribution is shown in green, highlighting that our constraint on $E_{\rm K,iso}$ is data-driven rather than prior-driven.}
    \label{fig:corner_jet_epsBfree}
\end{figure*}

\begin{figure*}
    \centering
    \includegraphics[width=0.6\textwidth]{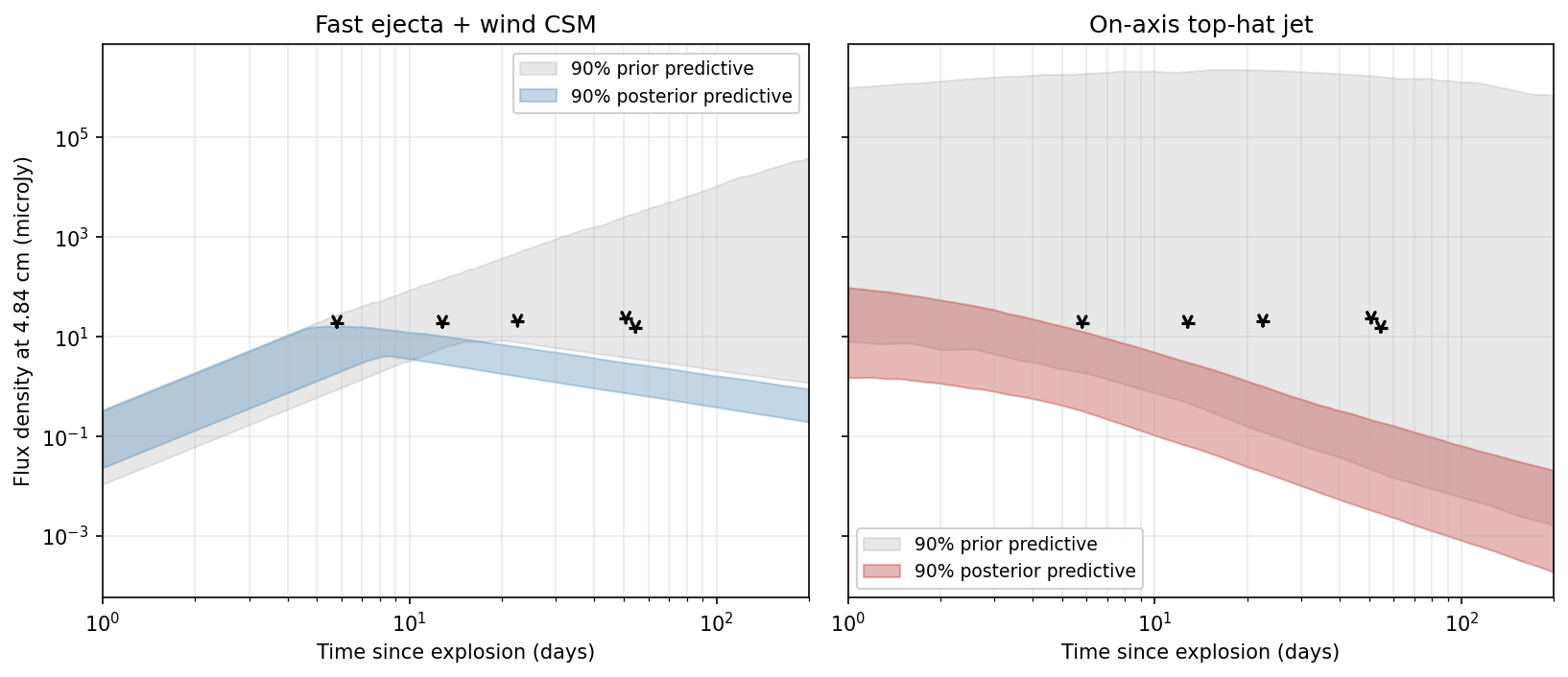}
    \caption{The allowed light curves for synchrotron radio emission from the fastest SN ejecta interacting with the outer, wind-like CSM (color) compared to the full light curve space probed by our priors. We fix $\epsilon_e = \epsilon_B = 0.1$.}
    \label{fig:corner_massloss}
\end{figure*}

\subsection{Host Galaxy SED Fitting}
\label{sec:host_appendix}

We obtain public photometric observations of the host using \texttt{FrankenBlast} \citep{frankenblast}, a customized version of the \texttt{Blast} web application \citep{blast}, that performs global aperture photometry on transient host galaxies using archival imaging catalogs including: the \textit{Galaxy Evolution Explorer} (GALEX; \citealt{Galex}), the Panoramic Survey Telescope and Rapid Response System (Pan-STARRS; \citealt{PanSTARRS}), Dark Energy Camera Legacy Survey Data Release 9 (DECaLS DR 9; \citealt{DECaLS}), the \textit{Two Micron All-Sky Survey} (2MASS; \citealt{Twomass}), and the \textit{Wide-field Infrared Survey Explorer} (WISE; \citealt{WISE}). We show the photometry of SDSS J095942.88+002506.2 used in our analysis in Table~\ref{tab:host}. We refer the reader to \citet{frankenblast} for specific details on the \texttt{FrankenBlast} aperture photometry technique. Additionally, we employ a spectrum of the host from the Dark Energy Spectroscopic Instrument (DESI) DR 1 \citep{desi_dr1} in our analysis. We observe prominent emission lines in the host spectrum, indicative of recent star formation: [O II]$\lambda3727$, H$\beta$, [O~III]$\lambda\lambda 4959, 5007$, H$\alpha$, and [S~II]$\lambda \lambda 6717, 6731$ at $z=0.0343$.

\begin{table}[ht]
    \centering
    \caption{Host Galaxy Photometry}
    \label{tab:host}
    \begin{tabular}{lc}
        \hline
        \hline
        Filter & AB mag \\
        \hline
\multicolumn{2}{c}{GLOBAL} \\
\hline
         GALEX FUV      & $19.45 \pm 0.11$ \\
         GALEX NUV      & $19.63 \pm 0.03$ \\
         DECaLS $g$      & $18.23 \pm 0.01$ \\
        DECaLS $r$      & $18.10 \pm 0.01$ \\
        DECaLS $z$      & $18.14 \pm 0.02$ \\
       Pan-STARRS $g$ & $18.36 \pm 0.01$ \\
       Pan-STARRS $r$ & $18.18 \pm 0.01$ \\
        Pan-STARRS $i$ & $18.18 \pm 0.02$ \\
        Pan-STARRS $z$ & $18.16 \pm 0.02$ \\
        Pan-STARRS $y$ & $18.21 \pm 0.05$ \\ 
        2MASS $J$      & $18.47 \pm 0.03$ \\
        WISE $w1$      & $19.50 \pm 0.17$ \\
        WISE $w2$      & $19.84 \pm 0.29$ \\
        WISE $w3$      & $17.61 \pm 0.58$ \\
        WISE $w4$      & $15.93 \pm 1.10$ \\
 \hline
\multicolumn{2}{c}{KNOT} \\
\hline
        DECaLS $g$ & $18.697 \pm 0.002$ \\
        DECaLS $r$ & $19.015 \pm 0.002$ \\
        DECaLS $i$ & $19.492 \pm 0.003$ \\
        DECaLS $z$ & $19.447 \pm 0.004$ \\
        \hline
    \end{tabular}

\tablecomments{Photometry of the host of SN\,2026gzf and the star-forming knot it is coincident with, used in the \texttt{Prospector} stellar population modeling fit. We note that the host was not detected in 2MASS $H$ and $K$, hence they were not included in fit.}
\end{table}

To obtain constraints on the global host galaxy stellar population properties (e.g., stellar mass, SFR), we model the host galaxy photometry and spectrum with \texttt{Prospector} \citep{Leja2019, jlc+2021}, a Python-based stellar population modeling inference code. We sample stellar population properties using a nested sampling fitting routine, \texttt{dynesty} \citep{Dynesty}, and build model SEDs with \texttt{FSPS} and \texttt{python-FSPS} \citep{FSPS_2009, FSPS_2010}. Internally, \texttt{Prospector} employs WMAP9 cosmology \citep{Hinshaw2013, Bennett+14}, \texttt{MIST} models \citep{MIST}, and the \texttt{MILES} spectral library \citep{MILES}. Our \texttt{Prospector} model includes the \citet{Chabrier2003}  initial mass function (IMF), a nebular emission model from \citet{bdc+2017}, and the \citet{KriekandConroy13} dust attenuation model, which measures an offset from the \citet{calzetti2000} attenuation curve and the amount of light attenuated from young ($\tau_{V,1}$) and old ($\tau_{V,2}$) stars. Given that the host is detected in multiple WISE filters, we also apply the \citet{DraineandLi07} IR dust emission model, in which we sample the polycyclic aromatic hydrocarbon mass fraction ($q_\textrm{pah}$), and a mid-IR AGN model, constrained through the mid-IR optical depth ($\tau_\textrm{AGN}$) and total AGN luminosity ($f_\textrm{AGN}$; a fraction of the total host bolomoteric luminosity). We sample stellar metallicity ($Z_*$) and total mass formed in the host ($M_F$) through the \citet{gcb+05} mass-metallicity relation, to ensure that we obtain a physically realistic $Z_*$, which can often become degenerate with other properties (e.g., dust and age) in SED fitting \citep{conroy2013}. We model the star formation history (SFH) in the host with a seven-bin non-parametric model, where we assume a constant SFR in each bin. The first two bins are spaced from 0-30~Myr and 30--100~Myr, and the last five bins are log-spaced until the age of the Universe at the redshift of the host.

To fit the host galaxy spectrum, we model the spectral continuum with a $6^\textrm{th}$ order Chebyshev polynomial, apply a model to normalize the spectral continuum to the observed photometry, and employ a spectral smoothing model that matches the resolution of the model spectrum to the observed spectrum. We additionally sample a gas-phase metallicity ($Z_\textrm{gas}$) and gas ionization parameter ($U_\textrm{gas}$) to fit the spectral lines. To ensure that the spectrum is not over-weighted in the fit in comparison to the photometry, we also include a model to inflate the spectral noise. Finally, we employ a pixel outlier model to marginalize over bad pixels, cosmic rays, and other sources of poorly-modeled noise in the spectrum. We show the \texttt{Prospector} SED fit (Section~\ref{sec:host_props}) of the host of SN 2026gzf in Figure~\ref{fig:host_sed}.

\begin{figure*}
    \centering
    \includegraphics[width=0.95\textwidth]{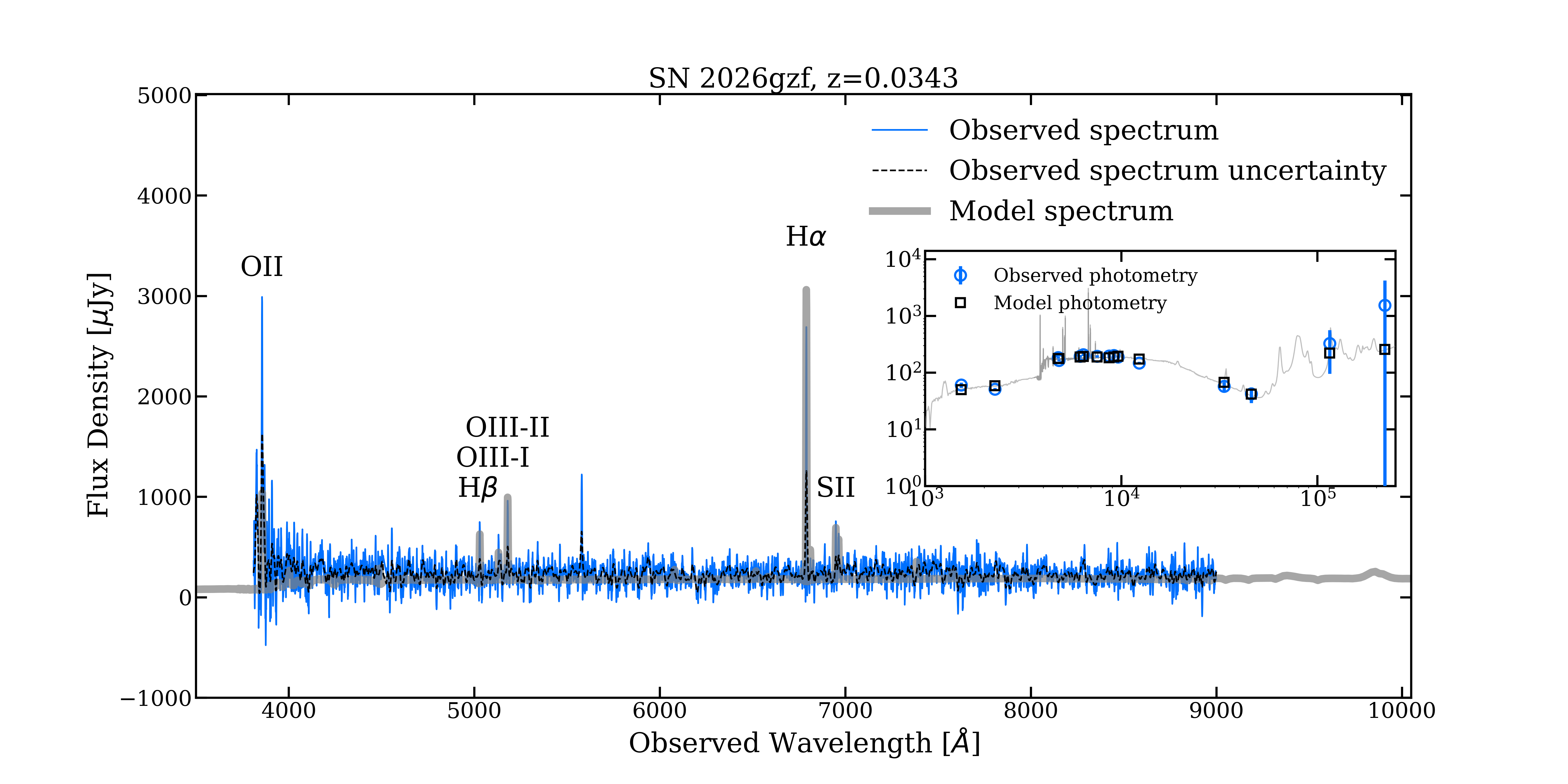}
    \caption{The \texttt{Prospector}-determined model spectrum (grey) compared to the the DESI-observed spectrum of the host of SN\,2026gzf (blue). We highlight high S/N lines in the observed spectrum. The inset shows the comparison between the model photometry (black squares) and observed photometry (blue circles). We find that \texttt{Prospector} fits the observed properties well, suggesting that the derived stellar population properties are accurate.}
    \label{fig:host_sed}
\end{figure*}

\subsection{Photometry Table}

We present the photometry shown in Figures~\ref{fig:opt_lc_spec} and \ref{fig:sne_models} in Table~\ref{tab:phot}.

\startlongtable
\begin{deluxetable*}{cccCc}
\tabletypesize{\footnotesize}
\centering
\tablecolumns{5}
\tabcolsep0.1in
\tablecaption{Photometric Observations of SN\,2026gzf
\label{tab:phot}}
\tablehead {
\colhead {Date}		&
\colhead {$\delta t$}		&
\colhead {Instrument}		&
\colhead {Filter}		& 
\colhead {Magnitude} \\
\colhead {} &
\colhead {(days)} &
\colhead {} &
\colhead {}	&
\colhead {(AB mag)} 
}
\startdata
61121.300 & 0.779 & ZTF & r & $19.85 \pm 0.18$ \\
61121.316 & 0.795 & ZTF & g & $19.65 \pm 0.10$ \\
61122.225 & 1.704 & ZTF & r & $18.84 \pm 0.04$ \\
61122.226 & 1.705 & ZTF & r & $18.79 \pm 0.09$ \\
61123.178 & 2.657 & WINTER & J & $18.78 \pm 0.11$ \\
61123.412 & 2.891 & SEDM & g & $17.90 \pm 0.05$ \\
61123.414 & 2.893 & SEDM & r & $18.07 \pm 0.04$ \\
61123.417 & 2.896 & SEDM & i & $18.26 \pm 0.04$ \\
61124.138 & 3.617 & SEDM & r & $17.75 \pm 0.04$ \\
61124.170 & 3.649 & SEDM & u & $18.14 \pm 0.05$ \\
61124.174 & 3.653 & SEDM & g & $17.78 \pm 0.04$ \\
61124.179 & 3.658 & SEDM & i & $17.89 \pm 0.04$ \\
61124.219 & 3.698 & SEDM & g & $17.76 \pm 0.03$ \\
61124.222 & 3.701 & SEDM & r & $17.78 \pm 0.03$ \\
61124.225 & 3.704 & SEDM & i & $17.91 \pm 0.04$ \\
61124.227 & 3.706 & SEDM & r & $17.62 \pm 0.19$ \\
61125.211 & 4.690 & ZTF & r & $17.55 \pm 0.05$ \\
61125.269 & 4.748 & ZTF & i & $17.79 \pm 0.04$ \\
61125.285 & 4.764 & SEDM & g & $17.58 \pm 0.04$ \\
61125.288 & 4.767 & SEDM & r & $17.48 \pm 0.03$ \\
61125.290 & 4.769 & SEDM & i & $17.64 \pm 0.03$ \\
61125.322 & 4.801 & ZTF & g & $17.54 \pm 0.03$ \\
61126.122 & 5.601 & WINTER & J & $18.24 \pm 0.14$ \\
61126.162 & 5.641 & ZTF & r & $17.36 \pm 0.02$ \\
61126.178 & 5.657 & SEDM & r & $17.33 \pm 0.04$ \\
61126.195 & 5.674 & ZTF & g & $17.27 \pm 0.01$ \\
61126.210 & 5.689 & SEDM & g & $17.40 \pm 0.02$ \\
61126.213 & 5.692 & SEDM & r & $17.30 \pm 0.04$ \\
61126.216 & 5.695 & SEDM & i & $17.49 \pm 0.04$ \\
61126.245 & 5.724 & ZTF & i & $17.54 \pm 0.03$ \\
61126.299 & 5.778 & SEDM & g & $17.36 \pm 0.04$ \\
61126.301 & 5.780 & SEDM & r & $17.29 \pm 0.04$ \\
61126.303 & 5.782 & SEDM & i & $17.46 \pm 0.04$ \\
61128.138 & 7.617 & SEDM & g & $17.22 \pm 0.70$ \\
61128.141 & 7.620 & SEDM & i & $17.06 \pm 0.14$ \\
61132.199 & 11.678 & ZTF & i & $17.16 \pm 0.04$ \\
61132.234 & 11.713 & ZTF & r & $16.96 \pm 0.06$ \\
61135.168 & 14.647 & ZTF & r & $16.92 \pm 0.04$ \\
61136.020 & 15.499 & LSST & r & $16.95 \pm 0.01$ \\
61136.023 & 15.502 & LSST & z & $17.04 \pm 0.01$ \\
61136.155 & 15.634 & LSST & i & $17.16 \pm 0.01$ \\
61137.144 & 16.623 & LSST & i & $17.15 \pm 0.01$ \\
61137.163 & 16.642 & ZTF & i & $17.14 \pm 0.08$ \\
61137.170 & 16.649 & ZTF & r & $16.93 \pm 0.05$ \\
61137.348 & 16.827 & SEDM & g & $17.47 \pm 0.11$ \\
61137.350 & 16.829 & SEDM & r & $16.92 \pm 0.04$ \\
61137.351 & 16.830 & SEDM & i & $17.08 \pm 0.03$ \\
61138.243 & 17.722 & WINTER & J & $17.51 \pm 0.10$ \\
61139.071 & 18.550 & LSST & g & $17.59 \pm 0.01$ \\
61139.073 & 18.552 & LSST & i & $17.19 \pm 0.01$ \\
61139.170 & 18.649 & ZTF & r & $16.93 \pm 0.06$ \\
61139.212 & 18.691 & ZTF & g & $17.58 \pm 0.07$ \\
61140.207 & 19.685 & ZTF & i & $17.15 \pm 0.02$ \\
61140.211 & 19.690 & ZTF & g & $17.68 \pm 0.06$ \\
61141.075 & 20.554 & LSST & i & $17.20 \pm 0.01$ \\
61141.125 & 20.604 & LSST & g & $17.79 \pm 0.01$ \\
61141.172 & 20.651 & LSST & i & $17.18 \pm 0.01$ \\
61142.065 & 21.544 & LSST & u & $19.58 \pm 0.01$ \\
61142.070 & 21.549 & LSST & z & $17.09 \pm 0.00$ \\
61143.003 & 22.482 & LSST & u & $19.74 \pm 0.02$ \\
61143.057 & 22.536 & LSST & i & $17.23 \pm 0.00$ \\
61143.059 & 22.538 & LSST & u & $19.74 \pm 0.02$ \\
61144.063 & 23.542 & LSST & u & $19.85 \pm 0.02$ \\
61145.126 & 24.605 & LSST & i & $17.28 \pm 0.01$ \\
61145.368 & 24.847 & SEDM & g & $18.33 \pm 0.34$ \\
61145.370 & 24.849 & SEDM & r & $17.20 \pm 0.04$ \\
61145.372 & 24.851 & SEDM & i & $17.11 \pm 0.07$ \\
61146.184 & 25.663 & ZTF & i & $17.33 \pm 0.06$ \\
61146.190 & 25.669 & ZTF & r & $17.20 \pm 0.06$ \\
61146.250 & 25.729 & SEDM & g & $18.41 \pm 0.04$ \\
61146.252 & 25.731 & SEDM & r & $17.23 \pm 0.04$ \\
61146.254 & 25.733 & SEDM & i & $17.18 \pm 0.04$ \\
61146.282 & 25.760 & ZTF & g & $18.39 \pm 0.10$ \\
61147.212 & 26.691 & WINTER & J & $17.58 \pm 0.11$ \\
61147.281 & 26.760 & ZTF & r & $17.29 \pm 0.06$ \\
61149.284 & 28.763 & SEDM & g & $18.70 \pm 0.30$ \\
61149.285 & 28.764 & SEDM & r & $17.52 \pm 0.05$ \\
61150.154 & 29.633 & ZTF & g & $18.55 \pm 0.12$ \\
61150.170 & 29.649 & ZTF & r & $17.40 \pm 0.04$ \\
61150.859 & 30.338 & IOO & r & $17.55 \pm 0.02$ \\
61150.860 & 30.339 & IOO & i & $17.68 \pm 0.04$ \\
61150.861 & 30.340 & IOO & z & $17.37 \pm 0.06$ \\
61150.862 & 30.341 & IOO & g & $18.44 \pm 0.03$ \\
61151.028 & 30.507 & LSST & i & $17.56 \pm 0.01$ \\
61151.031 & 30.510 & LSST & u & $20.52 \pm 0.02$ \\
61151.031 & 30.510 & LSST & u & $20.47 \pm 0.02$ \\
61151.289 & 30.768 & ZTF & g & $18.58 \pm 0.13$ \\
61152.083 & 31.562 & LSST & r & $17.57 \pm 0.01$ \\
61153.870 & 33.349 & IOO & r & $17.68 \pm 0.03$ \\
61153.871 & 33.350 & IOO & i & $17.75 \pm 0.05$ \\
61153.872 & 33.351 & IOO & z & $17.63 \pm 0.05$ \\
61153.874 & 33.353 & IOO & g & $18.66 \pm 0.05$ \\
61154.162 & 33.641 & ZTF & i & $17.60 \pm 0.03$ \\
61154.277 & 33.756 & SEDM & g & $18.99 \pm 0.26$ \\
61154.279 & 33.758 & SEDM & r & $17.62 \pm 0.04$ \\
61154.281 & 33.760 & SEDM & i & $17.54 \pm 0.04$ \\
61155.185 & 34.664 & ZTF & g & $18.91 \pm 0.16$ \\
61155.214 & 34.693 & ZTF & r & $17.84 \pm 0.07$ \\
61159.184 & 38.663 & ZTF & i & $17.91 \pm 0.05$ \\
61160.154 & 39.632 & SEDM & g & $19.11 \pm 0.05$ \\
61160.158 & 39.637 & SEDM & r & $17.86 \pm 0.04$ \\
61160.162 & 39.641 & SEDM & i & $17.76 \pm 0.05$ \\
61160.170 & 39.649 & ZTF & i & $17.93 \pm 0.09$ \\
61160.276 & 39.755 & ZTF & r & $17.98 \pm 0.13$ \\
61160.276 & 39.755 & ZTF & r & $18.12 \pm 0.09$ \\
61162.173 & 41.652 & ZTF & g & $19.14 \pm 0.10$ \\
61162.271 & 41.750 & ZTF & r & $18.02 \pm 0.12$ \\
61163.168 & 42.647 & ZTF & g & $19.11 \pm 0.12$ \\
61163.866 & 43.345 & IOO & r & $18.12 \pm 0.04$ \\
61163.867 & 43.346 & IOO & i & $18.16 \pm 0.09$ \\
61163.869 & 43.348 & IOO & z & $17.73 \pm 0.09$ \\
61163.870 & 43.349 & IOO & g & $19.04 \pm 0.04$ \\
61167.164 & 46.643 & ZTF & i & $18.28 \pm 0.09$ \\
61167.205 & 46.684 & ZTF & r & $18.15 \pm 0.08$ \\
61167.256 & 46.734 & ZTF & g & $19.27 \pm 0.16$ \\
61168.172 & 47.651 & ZTF & r & $18.22 \pm 0.09$ \\
61168.194 & 47.673 & ZTF & g & $19.36 \pm 0.16$ \\
61170.185 & 49.664 & ZTF & r & $18.31 \pm 0.07$ \\
\enddata
\tablecomments{
Observations are not corrected for Galactic nor local extinction. \\ Times are presented in the observer frame. }
\end{deluxetable*}

\begin{figure*}
    \centering
    \includegraphics[width=0.75\linewidth]{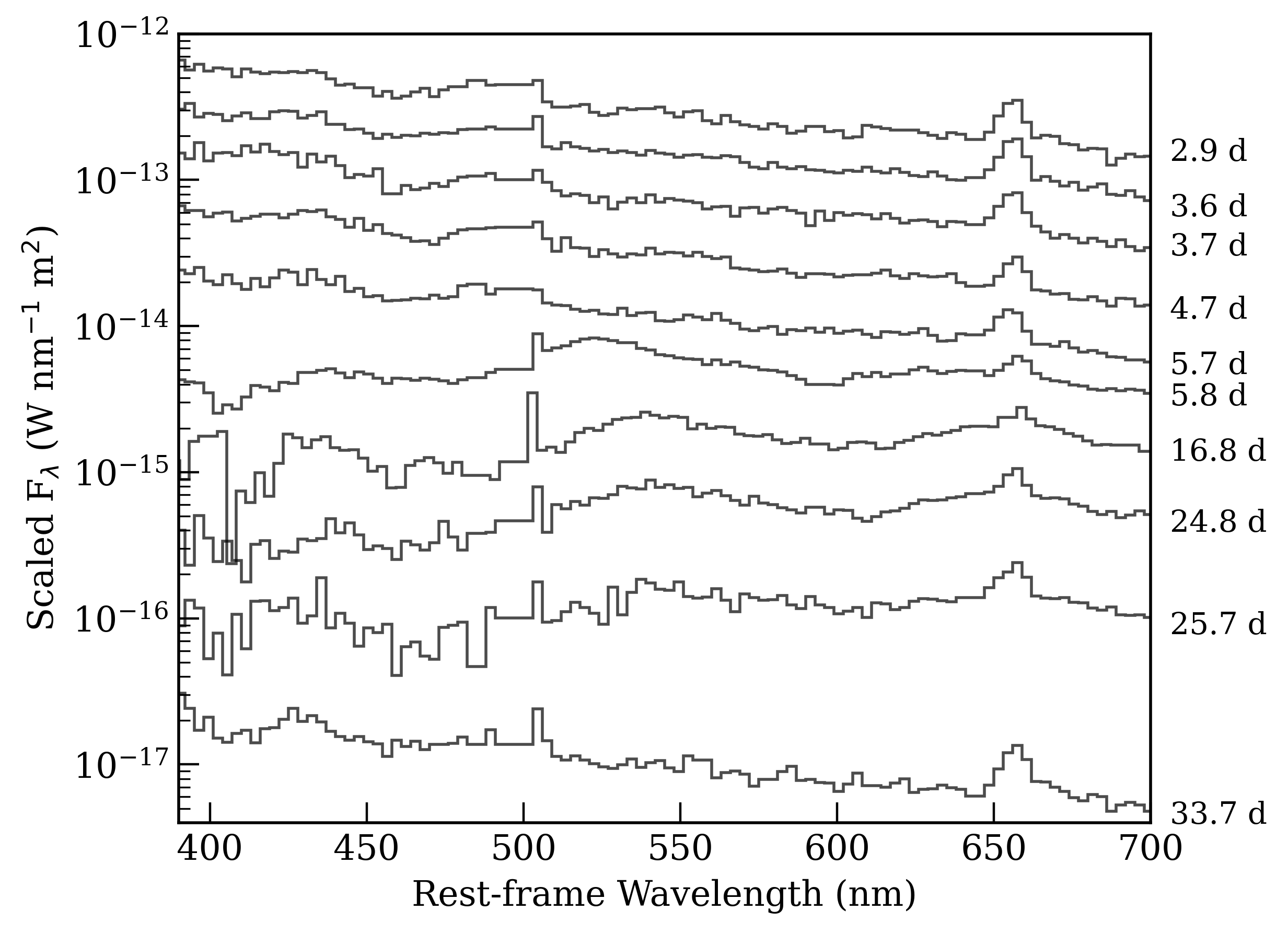}
    \caption{Spectral sequence of EP260321 from SEDM. Some narrow host galaxy lines are clipped for display purposes, and spectra are binned to 3 nm.}
    \label{sedmspec}
\end{figure*}

\end{document}